\documentclass[preprint2]{aastex631}
\usepackage{url,graphicx,cases}
\def\mbf#1{\mbox{\boldmath ${#1}$}}

\newcommand{\rev}[1]{#1}

\begin{document}

\title{The Growth of Protoplanets via the Accretion of Small Bodies in Disks Perturbed by the Planetary Gravity}
\author{Tatsuya Okamura}
\affiliation{Department of Physics, Nagoya University, Nagoya, Aichi 464-8602, Japan}
\email{okamura.tatsuya@nagoya-u.jp}
\author{Hiroshi Kobayashi}
\affiliation{Department of Physics, Nagoya University, Nagoya, Aichi 464-8602, Japan}
\begin{abstract}
Planets grow via the collisional accretion of small bodies in a protoplanetary disk.
Such small bodies feel strong gas drag and their orbits are significantly affected by the gas flow and atmospheric structure around the planet.
We investigate the gas flow in the protoplanetary disk perturbed by the gravity of the planet by three-dimensional hydrodynamic simulation.
We then calculate the orbital evolutions of particles in the gas structure obtained from the hydrodynamic simulation.
Based on the orbital calculations, we obtain the collision rate between the planet and centimeter to kilometer sized particles.
Our results show that meter-sized or larger particles effectively collide with the planet due to the atmospheric gas drag, which significantly enhances the collision rate.
On the other hand, the gas flow plays an important role for smaller particles.
Finally, considering the effects of the atmosphere and gas flow, we derive the new analytic formula for the collision rate, which is in good agreement with our simulations.
We estimate the growth timescale and accretion efficiency of drifting bodies for the formation of a gas-giant solid core using the formula.
We find the accretion of sub-kilometer sized bodies achieve \rev{a short growth timescale ($\sim 0.05\ {\rm Myr}$) and a high accretion efficiency ($\sim 1$) for the core formation at 5\,au in the minimum mass solar nebula model}.
\end{abstract}


\section{Introduction}
In the last ten years, the growth of protoplanets is considered via the accretion of mm-cm sized particles (pebbles) as well as km sized planetesimals \citep[e.g.,][]{ormel2010, lambrechts2012}. 
Pebbles, formed from dust grain coagulation, drift inward, and are effectively accreted onto protoplanets in inner protoplanetary disks.
\rev{These pebbles are aerodynamically small and well coupled to the gas, so that the gas flow significantly affects the collision rate between pebbles and planets.}
\rev{Recently, the detailed flow structures around a low-mass and non-gap-opening planet embedded in a protoplanetary disk were revealed by hydrodynamic simulations \citep{ormel2015b, fung2015, lambrechts2017, cimerman2017, kurokawa2018, kuwahara2019, bethune2019, fung2019, moldenhauer2021}}
The common flow structures of these studies are the horseshoe and vertical flow.
The horseshoe flow extends along the orbital direction of the planet in the anterior-posterior direction of the planet and has a vertical structure like a column.
The vertical flow comes from high altitudes to planets. 
These flow structures influence the collision rate between pebbles and the planet \rev{\citep{ormel2013}}.
\citeauthor{popovas2018}(\citeyear{popovas2018}, \citeyear{popovas2019}) showed the horseshoe flow takes pebbles away from the planet and prevents the particles from accreting onto the planet.
\cite{kuwahara2020a} showed the vertical flow helps small particles accreting onto the planet, so that the collision rate increases if pebbles have a vertical distribution. 
\rev{The analytic formula for the collision rates with planets has been derived for particles well coupled to the gas \citep{ormel2010,ormel2013}.}

On the other hand, the planetary atmosphere enhances the accretion rate of meter-sized or larger bodies.
Once protoplanets grow larger than the Moon, \rev{they can have atmospheres} \citep[e.g.,][]{mizuno1978}.
The gas density of the atmosphere may be \rev{in orders of magnitude larger} than the protoplanetary disk.
In addition, a close encounter with a planet accelerates the velocity of particles and the velocity may exceed the speed of sound.
These effects effectively reduce the kinetic energy of particles so that the particles are captured in the atmosphere.
The effective capture radius for sub-kilometer sized bodies is much larger than the physical radius of the planet \citep{inaba2003}.
\rev{For the accretion rate, \cite{inaba2003} derived the analytic formula, which is valid for accretion of sub-kilometer sized or larger particles. However, their formula overestimates the accretion rate of the meter-sized or smaller particles because the effective capture radius is estimated to be comparable to the Hill radius of the planet.}
\rev{Therefore, the atmosphere structure and flow around a protoplanet is necessary to derive accretion rates of centimeter to kilometer sized particles consistently.}

\cite{kurokawa2018} showed by hydrodynamic simulation, the atmospheric structure isolated from the protoplanetary disk is formed around the planet due to the cooling.
Therefore, the self-consistent flow and atmosphere around a planet is obtainable via hydrodynamic simulation. 
The accretion rate of bodies from pebbles to planetesimals can be calculated via the orbital simulation of bodies based on the density and velocity profile in the disk obtained via the hydrodynamic simulation.
\rev{Such combined simulations were carried out only for meter-sized or smaller particles \citep{popovas2018,popovas2019,kuwahara2020a,kuwahara2020b}. 
However, the accretion rate for a wide size range of particles is helpful to discuss the dominant accreted bodies onto a planet.}

In this paper, we investigate the collision rate between small bodies and a planet in the protoplanetary disk perturbed by the planetary gravity.
Our goal is to find out the effects of the gas flow and planetary atmosphere on the collision rate.
\rev{We perform hydrodynamic simulations and orbital calculations of centimeter to kilometer sized particles in the gas flow obtained from the hydrodynamic simulation.
Based on simulations, we derive the new analytic formula for the collision rate in the disk perturbed by the planetary gravity including the gas flow and planetary atmosphere.}
In \S \ref{sec:method}, we describe the simulation methods for hydrodynamics and orbital evolution of particles.
In \S \ref{sec:result}, we show the gas flow and atmospheric structure around a planet given by the hydrodynamic simulation and obtain the collision rate between the planet and particles via orbital calculation.
In \S \ref{sec:resultana}, we derive the new analytic formulae for the collision rate.
In \S \ref{sec:dis} and \ref{sec:con}, we discuss and summarize our findings.

\section{Methods}\label{sec:method}
\subsection{Hydrodynamic Simulations}

\subsubsection{Basic Equations}
We investigate the gas flow around a protoplanet with a circular orbit of radius $a$ in a protoplanetary disk around the host star with mass $M_{*}$ via hydrodynamic simulation.
We assume a compressible, inviscid, non-isothermal, non-self-gravitating fluid.
The basic equations of the hydrodynamic simulations are the equation of continuity, the Euler's equation, and the energy conservation equation, given by
\begin{equation}
\frac{\partial \rho_\mathrm{g}}{\partial t}+\nabla \cdot (\rho_\mathrm{g} \mbf{v}_\mathrm{g})=0,
\label{eq:eoc}
\end{equation}

\begin{equation}
\left( \frac{\partial}{\partial t }+ \mbf{v}_\mathrm{g} \cdot \nabla \right) \mbf{v}_\mathrm{g}=-\frac{\nabla p}{\rho_\mathrm{g}} +(\mbf{F}_\mathrm{cor}+\mbf{F}_\mathrm{tid}+\mbf{F}_\mathrm{p}),
\label{eq:euler}
\end{equation}

\begin{eqnarray}
\frac{\partial E}{\partial t}+\nabla \cdot [(E+p)\mbf{v}_\mathrm{g}]&=&\rho_\mathrm{g} \mbf{v}_\mathrm{g} \cdot(\mbf{F}_\mathrm{cor}+\mbf{F}_\mathrm{tid}+\mbf{F}_\mathrm{p}) \nonumber \\
&-&\frac{U(\rho_\mathrm{g},T)-U(\rho_\mathrm{g},T_0)}{\beta/\Omega},
\label{eq:energy}
\end{eqnarray}
where $\rho_\mathrm{g}$ is the gas density, $\mbf{v}_\mathrm{g}$ is the gas velocity, $p$ is the pressure, $\mbf{F}_\mathrm{cor}$ is the Coriolis force, $\mbf{F}_\mathrm{tid}$ is the tidal force, $\mbf{F}_\mathrm{p}$ is the gravitational force, $\Omega$ is the orbital frequency, $\beta$ is the dimensionless cooling timescale, $T$ and $T_0$ is the fluid and background temperatures, respectively, and $U$ and $E$ are, respectively, the internal and total energy densities, given by
\begin{equation}
U = \frac{p}{\gamma-1},
\end{equation}
\begin{equation}
E=U+\frac{1}{2}\rho_\mathrm{g} v_\mathrm{g}^{2},
\end{equation}
with the ratio of specific heat $\gamma=7/5$.

The forces in Eqs. (\ref{eq:euler}) and (\ref{eq:energy}) 
are given by $\mbf{F}_\mathrm{cor}=-2\Omega\mbf{e}_z\times\mbf{v}_\mathrm{g}$, $\mbf{F}_\mathrm{tid}=3 x \Omega^2 \mbf{e}_x - z \Omega^2 \mbf{e}_z$,
and
\begin{equation}
\mbf{F}_\mathrm{p}=\nabla\left(\frac{G M_\mathrm{p}}{\sqrt{r^2+r^2_\mathrm{s}}}\right)\left\{1-\mathrm{exp}\left[-\frac{1}{2}\left(\frac{t}{t_\mathrm{inj}}\right)^2\right]\right\},
\end{equation}
where $\mbf{e}_i$ is the unit vector in the $i$-direction, $M_\mathrm{p}$ is the mass of the planet, $G$ is the gravitational constant, $r$ is the distance from the center of the planet, $r_\mathrm{s}$ is the softening length, and $t_\mathrm{inj}$ is the injection time of planetary gravity.
To avoid the numerical effect, the gravity of the planet is gradually inserted using the injection time $t_\mathrm{inj}=0.5\Omega^{-1}$ \citep{ormel2015a}.
The planet have an atmosphere with radius $\sim R_{\rm B}$, where $R_{\rm B} \equiv G M_{\rm p}/c_{\rm s}^2$ is the Bondi radius and $c_{\rm s}$ is the isothermal sound speed.
To resolve the atmospheric structure, we set $r_\mathrm{s}=0.1 R_{\rm B}$ according to \cite{kurokawa2018}. 

The last term in Eq. (\ref{eq:energy}) is the cooling function according to the $\beta$ cooling model where the temperature $T$ relaxes toward $T_0$ in the timescale $\beta/\Omega$ \citep[e.g.,][]{gammie2001}.
In this study, we adopt the $\beta$ cooling model to obtain the density profile and flow around the planetary atmosphere according to \cite{kurokawa2018}, who showed the atmosphere is formed around the planet using the $\beta$ cooling model.
We set $\beta=(R_\mathrm{B} \Omega/0.1 c_{\rm s})^{2}$ according to the previous studies \citep{kurokawa2018, kuwahara2020a, kuwahara2020b}.

The time, velocity, and length are considered to be normalized by the reciprocal of the Keplerian orbital frequency $\Omega^{-1}$, the isothermal sound speed $c_\mathrm{s}$, and the gas scale hight $H\equiv c_\mathrm{s}/\Omega$, respectively.
The basic equations are then characterized by a dimensionless number, given by
\begin{equation}
m=\frac{R_\mathrm{B}}{H}=\frac{G M_\mathrm{p}\Omega}{c_\mathrm{s}^3}.
\end{equation}
The Hill radius is written as a function of $m$:
\begin{equation}
R_\mathrm{H}=\left(\frac{m}{3}\right)^{1/3} H.
\end{equation}
We assume a solar mass host star and a disk temperature profile $T=270(a/1\ \mathrm{au})^{-1/2}$\citep[i.e., the minimum mass solar nebula model,][]{weidenschilling1977b, hayashi1985}, so that $M_\mathrm{p}$ is given by \citep{kurokawa2018}
\begin{equation}
M_\mathrm{p} \simeq 12m\left(\frac{a}{1\ \mathrm{au}}\right)^{3/4} M_\oplus .
\end{equation}
We investigate the cases with $m=0.1$, $0.05$, and $0.03$, which correspond to $M_{\rm p}/M_{\oplus} = 1.2$ $(4.0)$, $0.6$ $(2.0)$, and $0.36$ $(1.2)$ at $1$ $(5)$ au, respectively.

\subsubsection{Simulation Setups and Boundary Conditions}

In this study, we use the hydrodynamic simulation code Athena++ \citep{white2016, stone2020}.
We choose the HLLC algorithm for a Riemann solver.
Our simulations are performed in a spherical-polar coordinate $(r,\theta,\phi)$ centered on the planet.
\rev{We focus on the embedded, non-gap-opening regime and thus the local simulations are likely to be valid.}
In order to resolve the flow structure around the planet in detail, we adopt a logarithmic grid for the radial dimension.
In the polar angle direction, the cell size is proportional to $(3\psi^2+1)$, where $\psi$ is the angle from the midplane.
The cell spacing near the midplane is smaller than near the pole \citep[i.e., resolution near the midplane is higher than near the pole;][]{kurokawa2018}.
The numerical resolution is set to be $128\times 64 \times128$ in the $r$, $\theta$, and $\phi$ directions, respectively.
The computational domain ranges from $0$ to $2\pi$ for $\phi$, from $0$ to $\pi$ for $\theta$, and from $r_\mathrm{inn}$ to $r_\mathrm{out}$ for $r$, where $r_\mathrm{inn}$ and $r_\mathrm{out}$ are the radii at the inner and outer boundaries, respectively.
\par
We assume the initial and outer boundary values in hydrodynamic simulations are those of a Kepler disk without pressure gradient term, therefore given by

\begin{equation}
\rho_{\mathrm{g},0}=\rho_0\ \mathrm{exp}\left[-\frac{1}{2}\left(\frac{z}{H}\right)^2\right],
\label{eq:initialdensity}
\end{equation}

\begin{equation}
\mbf{v}_{\mathrm{g},0}=-\frac{3}{2}  \Omega x \mbf{e}_y,
\label{eq:initialvelocity}
\end{equation}
where $\rho_0$ is the density at the midplane and $z$ is the distance from the midplane.
\rev{We focus on the shear regime of pebble accretion, where the shear velocity is more dominant than the headwind of the gas 
to determine the accretion rate.}
\rev{The condition satisfied to be in the shear regime is given by \citep{ormel2017}}
\begin{eqnarray}
M_\mathrm{p} &\gg& \frac{1}{8}\frac{v_\mathrm{hw}^3}{G \Omega St_0} \nonumber \\
 &\simeq& 2.0\times10^{-4}M_\oplus \frac{1}{St_0} \nonumber \\ 
 &\times&\left(\frac{a}{1\  \mathrm{au}}\right)^{3/2}\left(\frac{v_\mathrm{hw}}{50\ \mathrm{m\ s^{-1}}}\right)^{3},
\end{eqnarray}
\rev{where $v_{\rm hw}$ is the headwind of the gas and $St_0$ is the dimensionless stopping time of a particle (see Eq. \ref{eq:stradii}).}
\rev{We perform simulations for $M_\mathrm{p}=0.36-1.2\ M_\oplus$ at $a=1\ {\rm au}$ ($M_{\rm p}=1.2-4.0\ M_{\oplus}$ at $a=5\ {\rm au}$) and $St_0=3.0\times10^{-3}-1.0\times10^{13}$, so that we ignore the headwind in our simulations.}

We set the radius of the inner boundary according to that of the planet \citep{kuwahara2020a}.
We assume the density of the planet $\rho_\mathrm{pl}=5\ \mathrm{g}/\mathrm{cm}^{3}$, so that the radius of the inner boundary is given by 
\begin{eqnarray}
r_\mathrm{inn}&=&\left(\frac{3M_\mathrm{p}}{4 \pi \rho_\mathrm{pl}}\right)^{1/3} = \left(\frac{9M_{*}}{4 \pi \rho_\mathrm{pl}}\right)^{1/3} \frac{R_{\rm H}}{a} \nonumber  \\ 
&\simeq& 3\times10^{-3} m^{1/3} \nonumber \\
&\times& \left(\frac{\rho_\mathrm{pl}}{5\ \mathrm{g\ cm^{-3}}}\right)^{-1/3} \left(\frac{M_*}{1M_\odot}\right)^{1/3} \left(\frac{a}{1\ \mathrm{au}}\right)^{-1}. \label{eq:rplanet}
\end{eqnarray}

We introduce the reflective boundary condition for the inner boundary, but \cite{kurokawa2018} reported this condition generates the unphysical energy flow \rev{in the results of simulations by Athena++}.
In order to prevent this unphysical affair from affecting the entire flow, we introduce the artificial cooling at the three inner cells \citep[$\beta=10^{-5}$;][]{kurokawa2018}.
The boundary condition in the azimuthal direction is set to the periodic boundary, which means $A(r,\ \theta,\ \phi)=A(r,\ \theta,\ \phi+2\pi)$ holds for an arbitrary physical quantity $A$.
We calculate until the steady state flow is almost achieved at $t=t_{\rm end}$. We set $t_{\rm end}$ according to \cite{kuwahara2020a}.
A summary of our simulation parameters is showed in Table \ref{tab:hydropara}, which are chosen from the values at $a=1\ {\rm au}$.

\begin{table*}[htbp]
	\begin{center}
	\caption{List of parameters for our simulations. The first column shows the value of dimensionless mass of the planet. The second to eighth columns represent the corresponding values at $1\ {\rm au}$ for the mass of the planet normalized by the earth mass, the Bondi radius, the Hill radius, the size of the inner boundary, the size of the outer boundary, termination time of the hydrodynamic simulation, dimensionless cooling timescale $\beta$.
}
	\begin{tabular}{cccccccc} \hline
	m & physical mass ($M_{\oplus}$) & $R_\mathrm{B}\ (H)$ & $R_\mathrm{H}\ (H)$ & $r_\mathrm{inn}\ (H)$ &  $r_\mathrm{out}\ (H)$ & $t_\mathrm{end}\ (\Omega^{-1})$ & $\beta$ \\  \hline \hline 
	0.03 & 0.36 & 0.03 & 0.22 & $9.32\times10^{-4}$ & 0.5 & 50 & 0.09 \\
	0.05 & 0.6  & 0.05 & 0.26 &  $1.1\times10^{-3}$ & 0.75 & 100 & 0.25\\
	0.1 & 1.2 & 0.1 &  0.32 &  $1.39\times10^{-3}$ & 5 & 150 & 1
    	\end{tabular}
	\label{tab:hydropara}
	\end{center}
\end{table*}


\subsection{Orbital Calculations}
\subsubsection{Equations and Gas Drag Law}
We calculate \rev{orbits of }particles in the gas flow obtained from hydrodynamic simulations.
The equation of motion of the particle is given by
\begin{eqnarray}
\frac{d\mbf{v}}{dt}=
\left(
\begin{array}{ccc}
2 v_y \Omega + 3 x \Omega^2\\
-2 v_x \Omega\\
-z\Omega^2
\end{array}
\right)
-\frac{\mathrm{G}M_\mathrm{p}}{r^3}
\left(
\begin{array}{ccc}
x \\
y \\
z
\end{array}
\right)
+\frac{\mbf{F}_\mathrm{drag}}{m_\mathrm{p}}, \nonumber \\
\label{eq:orbiteq}
\end{eqnarray}
where $\mbf{v}=(v_x,v_y,v_z)$ is the velocity vector of the particle, $\mbf{r}=(x,y,z)$ is its position vector with respect to the center of the planet, and $m_\mathrm{p}$ is the mass of the particle.
The first and second terms on the right-hand side of Eq. (\ref{eq:orbiteq}) are the Coriolis and tidal forces and the gravity force of the planet, respectively.
\rev{The third term is the gas drag force given by \citep[e.g.,][]{adachi1976}}
\begin{equation}
\mbf{F}_\mathrm{drag}= -\frac{C_\mathrm{D}}{2} \pi r_\mathrm{p}^2 \rho_\mathrm{g} u \mbf{u},
\label{eq:fdrag}
\end{equation}
\rev{where $r_\mathrm{p}$ is a particle radius, $\mbf{u}\equiv \mbf{v}-\mbf{v}_\mathrm{g},\ u=|\mbf{u}|$ is the velocity of a particle relative to the gas, and $C_\mathrm{D}$ is the gas drag coefficient. }
\rev{We consider centimeter to kilometer sized particles; $C_\mathrm{D}$ is given by the Stokes gas drag for small particles, while $C_\mathrm{D}$ is constant for large particles \citep{adachi1976}.
In addition, if particle velocities exceed the sound speed due to planetary gravity, the supersonic gas drag law should be applied.
The gas drag coefficient, $C_\mathrm{D}$ is approximated to be \citep{adachi1976, tanigawa2014}.}
\begin{equation}
C_\mathrm{D} = \frac{12 \nu}{r_{\rm p} u} + \frac{(2-w)M}{1.6+M}+w,
\label{eq:cd}
\end{equation}
\rev{where $\nu$ is the kinetic viscosity, $M\equiv u/c_\mathrm{s}$ is the Mach number, and $w=0.4$ is the correction factor.}
\rev{In Eq. (\ref{eq:cd}), the first, second, and third terms indicate Stokes, quadratic, and supersonic drag coefficients, respectively, and we ignore Epstein drag for simplicity.}
In our simulation, we do not consider the evaporation or ablation.
The kinetic viscosity is given by $\nu=l_{\rm mfp} c_{\rm s}/2$, where $l_{\rm mfp}=\mu m_{\rm H}/\rho_{\rm g}\sigma$ is the mean free path of the gas, $\mu=2.34$ is the mean molecular weight, $m_{\rm H}=1.67\times 10^{-24}\ {\rm g}$ is the mass of the proton, and $\sigma=2\times 10^{-15}\ {\rm cm}^2$ is the molecular collision cross section \citep{chapman1970, adachi1976}.
The stopping time of a particle is expressed by
 \begin{equation}
t_\mathrm{stop}= \frac{m_\mathrm{p} u}{|\mbf{F}_\mathrm{drag}|}=\frac{4 \rho_\mathrm{p} r_\mathrm{p}^2 }{9 \rho_\mathrm{g} c_\mathrm{s} l_\mathrm{mfp}}\left(1+ \frac{2M+1.6w}{1.6+M} \cdot \frac{r_\mathrm{p}u}{6 c_\mathrm{s} l_\mathrm{mfp}}\right)^{-1},
\end{equation}
where $\rho_\mathrm{p}$ is the density of a particle.
We investigate the orbits of particles with different radii.
We introduce the Stokes parameter, $St$ defined as the stopping time multiplied by $\Omega$. The initial particle has $u\ll c_{\rm s}$ so that the gas drag is mainly given in the Stokes gas drag regime. This initial Stokes parameter is approximated to be
\begin{equation}
St_0=4.5 \times 10^{-4} \left(\frac{\rho_\mathrm{p}}{1\ \mathrm{g}\ \mathrm{cm}^{-3}}\right) \left(\frac{c_\mathrm{s}}{1\ \mathrm{km\ s^{-1}}}\right)^{-1} \left(\frac{r_\mathrm{p}}{1\ \mathrm{cm}}\right)^{2},
\label{eq:stradii}
\end{equation}
where the value of $c_\mathrm{s}$ is chosen as that approximately at $1\ \mathrm{au}$ in the minimum mass solar nebula model.
Note that $St$ continuously changes during the calculation.
The value of $St$ can be less than $St_0$ due to the gas drag for $u \gg c_{\rm s}$.
We use $St_0$ given in Eq. (\ref{eq:stradii}) as a parameter instead of the radii of particles.


\subsubsection{Numerical Setups: 3D Orbital Calculations}
\par
A particle is launched from a starting point ($x_\mathrm{s},y_\mathrm{s},z_\mathrm{s}$), where $y_\mathrm{s}$ is fixed at $y_\mathrm{s}=40R_\mathrm{H}$, and $x_\mathrm{s}$ and $z_\mathrm{s}$ are varied \citep{ida1989, ormel2010}.
The particle is initially well coupled to the gas for $St_0 \ll 1$, while the motion is determined by Kepler's laws for $St_0 \gg 1$.
We thus set the initial condition according to the gas motion for $St_0 <1$ or the orbital elements for $St_0 > 1$. 
If the orbital eccentricities of particles are much smaller than the Hill radius of the planet divided by $a$, the collision rate is independent of eccentricities \citep{ida1989}.
We set initially circular orbits.
The initial conditions are given as follows.
\begin{eqnarray}
z_{\rm s}&=&
\left\{
\begin{array}{cc}
z_0\ \ \ (St_0<1),\\
i a_{\rm s}\ \mathrm{sin}\ \omega \equiv z_0\ \mathrm{sin}\ \omega \ \ \  (St_0>1), \nonumber \\
\end{array}
\right. \\
\mbf{v}_0 &=& \left( 0, -\frac{3}{2}\Omega x_\mathrm{s}, v_{z,0} \right), \label{eq:initialz} \\ 
v_{z,0}&=&
\left\{
\begin{array}{cc}
0\ \ \ (St_0<1),\\
z_0\ \mathrm{cos}\ \omega  \ \ \  (St_0>1), \nonumber \\
\end{array}
\right.
\end{eqnarray}
where $i$ is the initial inclination of the particle, $a_{\rm s}$ is its initial semimajor axis and $\omega$ is its longitude of ascending node.
We assume $\omega$ is distributed uniformly in the range between 0 and $\pi$, and take the average for the calculation of the collision rate.
\rev{For $St_0<1$, we ignore the $z$-component of the tidal force in Eq. (\ref{eq:orbiteq}) assuming the balance with turbulent stirring according to \citet{kuwahara2020a, kuwahara2020b}.}

The lower limits of $x_\mathrm{s}$ and $z_\mathrm{0}$ are set to $x_\mathrm{s}=0.0001H$ and $z_\mathrm{0}=0$ and the spacial intervals of $x_\mathrm{s}$ and $z_\mathrm{0}$ are $0.0001H$ and $0.02H$.
The upper limits of $x_\mathrm{s}$ and $z_\mathrm{0}$ are set to a few disk scale hights.
We calculate orbits only coming from positive $x$ and $y$ because of the symmetry.

We use $\rho_\mathrm{g}$, $\mbf{v}_\mathrm{g}$, and $c_\mathrm{s}$ given from the hydrodynamic simulation at $t=t_\mathrm{end}$, at which the fluid is in a quasi-steady state.
The starting point of the particle is out of the computational domain of our hydrodynamic simulation ($r_\mathrm{out}$), so that we set the velocity and density of gas in $r > r_\mathrm{out}$ are the same as the outer boundary conditions given in Eqs. (\ref{eq:initialdensity}) and (\ref{eq:initialvelocity}).
In our hydrodynamic simulations  for $m=0.03$ and $m=0.05$, unexpected vortices are appeared in the horseshoe structure.
These vortices are found by \citet{kuwahara2020a, kuwahara2020b}, which are caused by the low resolution of the distant place from the planet.
In these cases, we only use the results of the hydrodynamic simulation in $r<0.3H\ (m=0.03)$ and in $r<0.55H\ (m=0.05)$, respectively.
All physical quantities obtained via hydrodynamic simulations are the discrete data, so that we interpolate physical quantities using the linear interpolation method \citep{kuwahara2020a}.

Orbital integration is terminated if any one of the following conditions is satisfied.
(i) A particle goes far away; $|y|>40R_\mathrm{H}$. (ii) A particle collides with the planet; $r<r_\mathrm{inn}$, where $r_\mathrm{inn}$ is the inner boundary for hydrodynamic simulations and the physical radius of the planet in Eq. (\ref{eq:rplanet}).

We numerically integrate Eq. (\ref{eq:orbiteq}) using the Runge-Kutta-Fehlberg scheme \citep{fehlberg1969, eshagh2005, ormel2010}.
This integration method controls each time step comparing a fourth-order solution with a fifth-order solution. We set the relative error tolerance of $10^{-8}$, which ensures numerical convergence \citep{ormel2010, visser2016}.

\begin{figure}[htb!]
	\centering
	\includegraphics[scale=0.45]{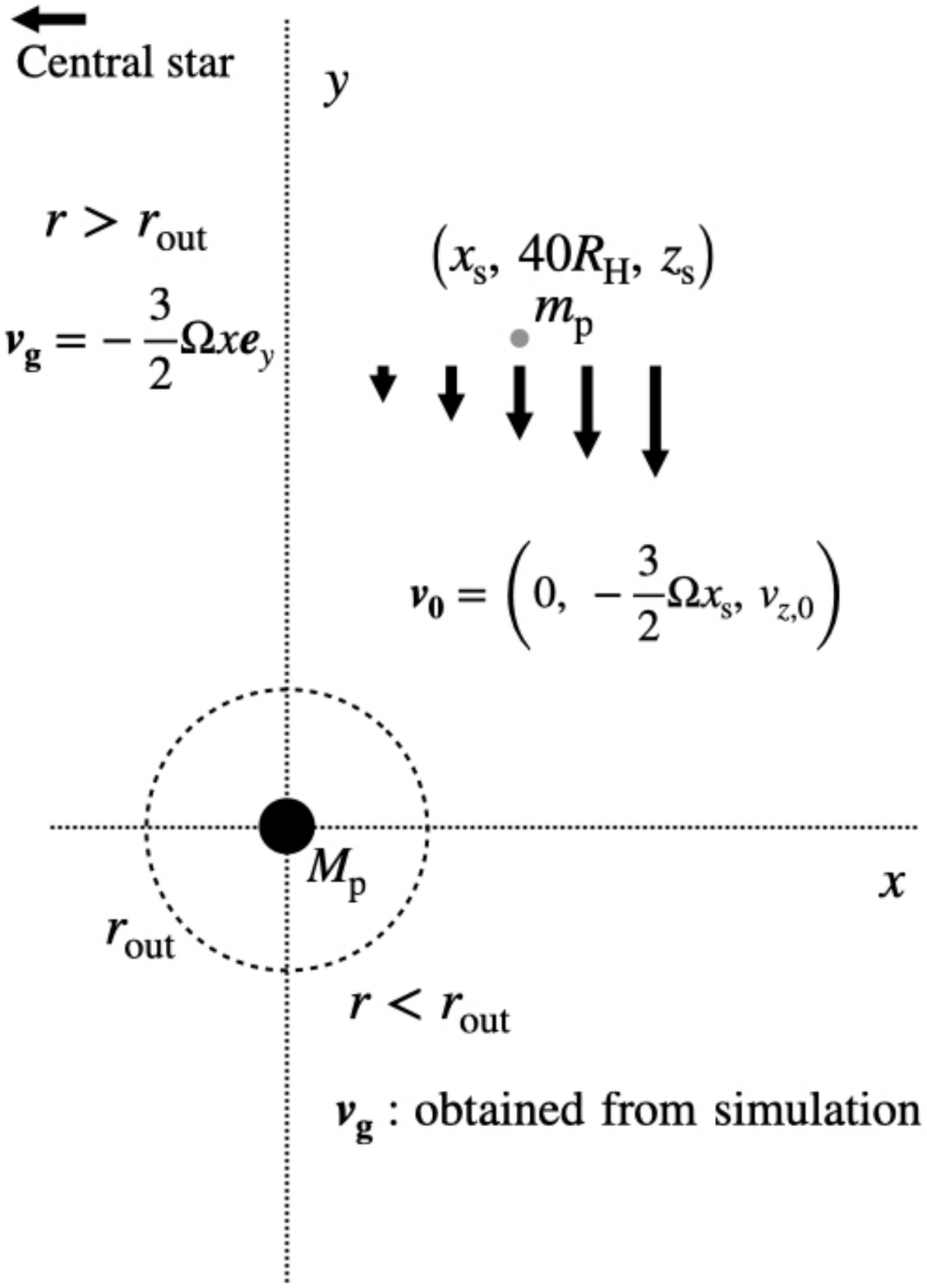}
	\caption{Sketch of the orbital calculation of particles in the co-rotating frame. A planet is located at the origin of the frame. We calculate orbits only coming from positive $x$ and $y$ because of the symmetry. The dashed circle shows the outer radius of the hydrodynamic simulation. If the particle is within the radius, we use the gas velocity and density obtained from the hydrodynamic simulation. On the other hand, the particle is out of the radius, we use the gas velocity and density that is the same as the outer boundary conditions of the hydrodynamic simulation.}
	\label{fig:image}
\end{figure}

We show the sketch of the orbital calculation of particles in Fig. \ref{fig:image}.

\section{Results of Simulations}\label{sec:result}
\subsection{Hydrodynamic Simulation}
\begin{figure}[htb!]
	\centering
	\includegraphics[scale=0.3]{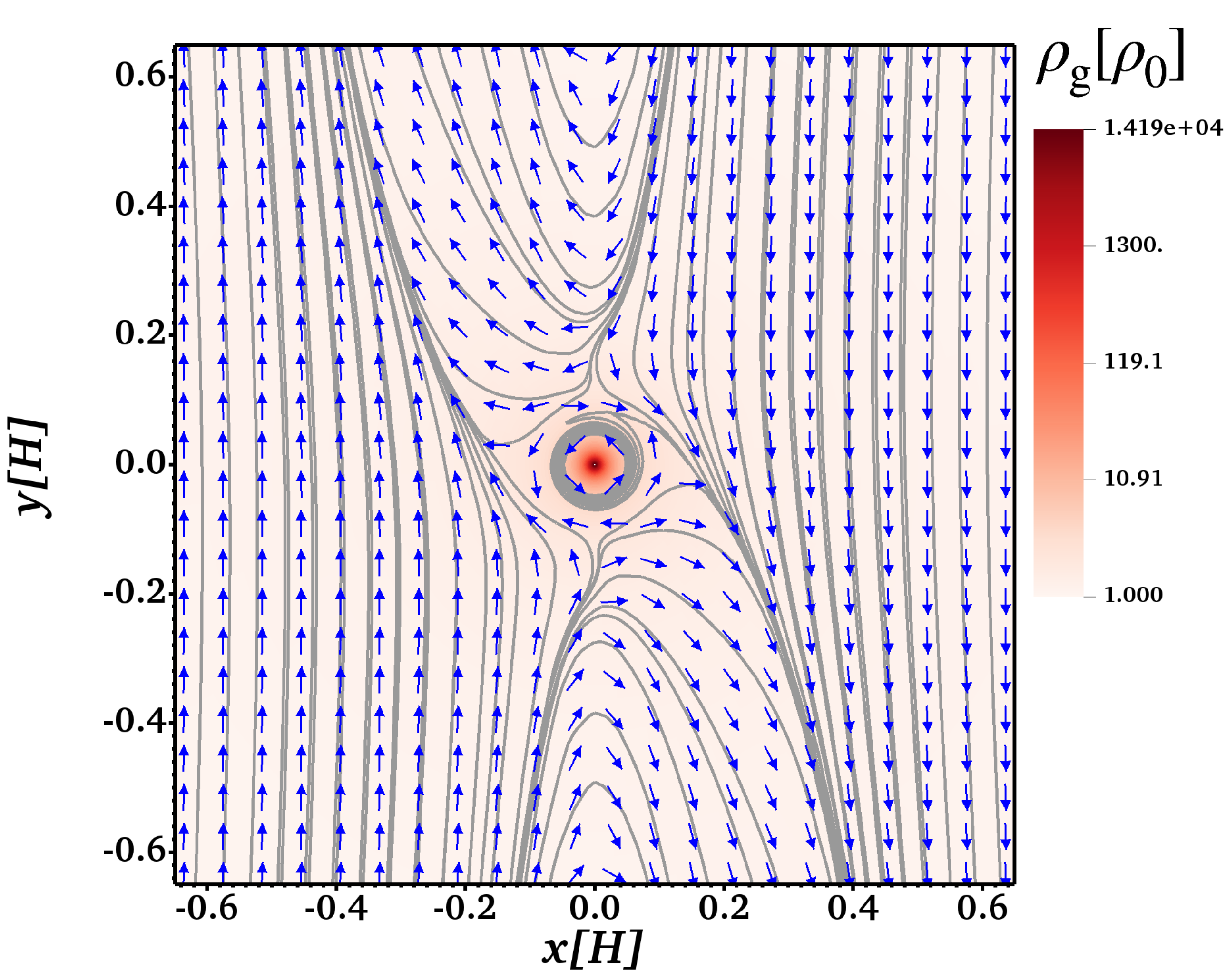}
	\caption{Flow patterns in the midplane ($z=0$) for $m=0.1$ at $t=150\ \Omega^{-1}$ are shown by blue arrows and streamlines. The length of the arrows do not scale to the gas velocity. Color contour shows the gas density.}
	\label{fig:stream}
\end{figure}

\begin{figure}[htb!]
	\centering
	\includegraphics[scale=0.3]{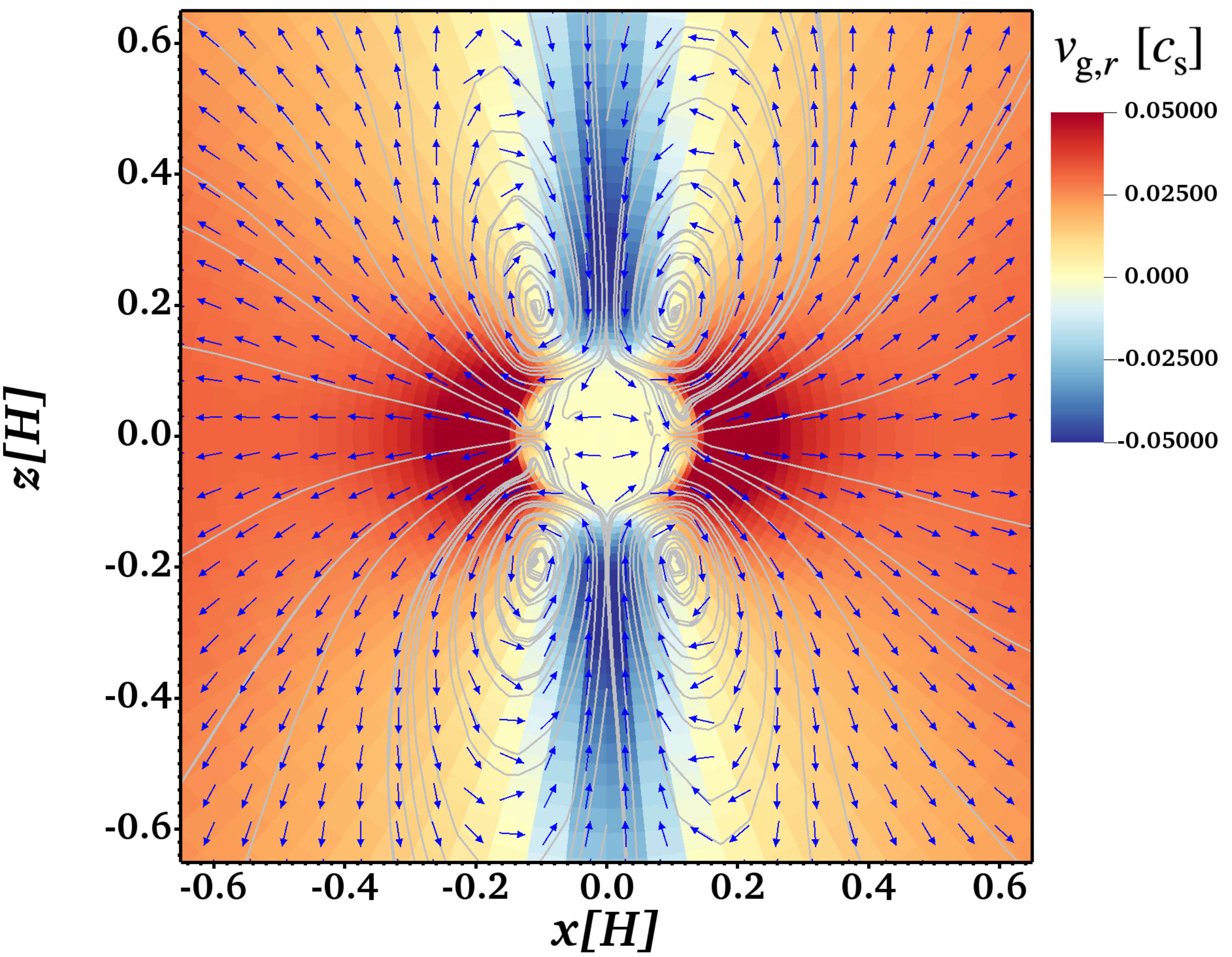}
	\caption{Flow patterns in $x$-$z$ plane ($y=0$) for $m=0.1$ at $t=150\ \Omega^{-1}$ are shown by blue arrows and streamlines. The length of the arrows do not scale to the gas velocity. Color contour shows the velocity in the radial direction.}
	\label{fig:vz}
\end{figure}

Figures \ref{fig:stream} and \ref{fig:vz} show the flow structures in the midplane ($z=0$) and in the plane of $y=0$ at $t=t_\mathrm{end}$, respectively.
The additional simulations for $t>t_\mathrm{end}$ show that the flow structure and physical quantities almost keep constant subsequent to $t=t_\mathrm{end}$, so that the fluid is in a quasi-steady state.
The characteristic structures of the gas flow formed in the simulations are similar to the previous studies \citep{ormel2015b, fung2015, lambrechts2017, cimerman2017, kurokawa2018, popovas2018, kuwahara2019, chrenko2019, bethune2019, fung2019}.
These structures are divided into four parts.  
(i) The Keplerian shear streams exist the distant place from the planet ($|x| \gtrsim 0.4H$ in Fig. \ref{fig:stream}). 
These streamlines are almost same as the initial and outer boundary conditions, because the planet is too far to change the flow. 
(ii) The horseshoe flow extends along the orbital direction of the planet in the anterior-posterior direction of the planet.
The U-turn flows caused by the horseshoe flow are seen in $|x|\lesssim 0.1 H$ and $|y| \gtrsim 0.2 H$ in Fig. \ref{fig:stream}.
(iii)  The vertical flow comes from high altitudes to planets, as seen in $|x| \lesssim 0.1 H$ and $|z| \gtrsim 0.1 H$ in Fig. \ref{fig:vz}.
Of course, this result is specific to three-dimensional calculations.
(iv) The atmospheric structure is formed around the planet, as seen in $|r| \lesssim 0.1H$ in Figs. \ref{fig:stream} and \ref{fig:vz}.
This region is isolated from the outer flow, which means the gas recycling does not occur.
This isolated envelope is formed due to the cooling \citep{kurokawa2018}.

\begin{figure}[htb!]
	\centering
	\includegraphics[scale=0.3]{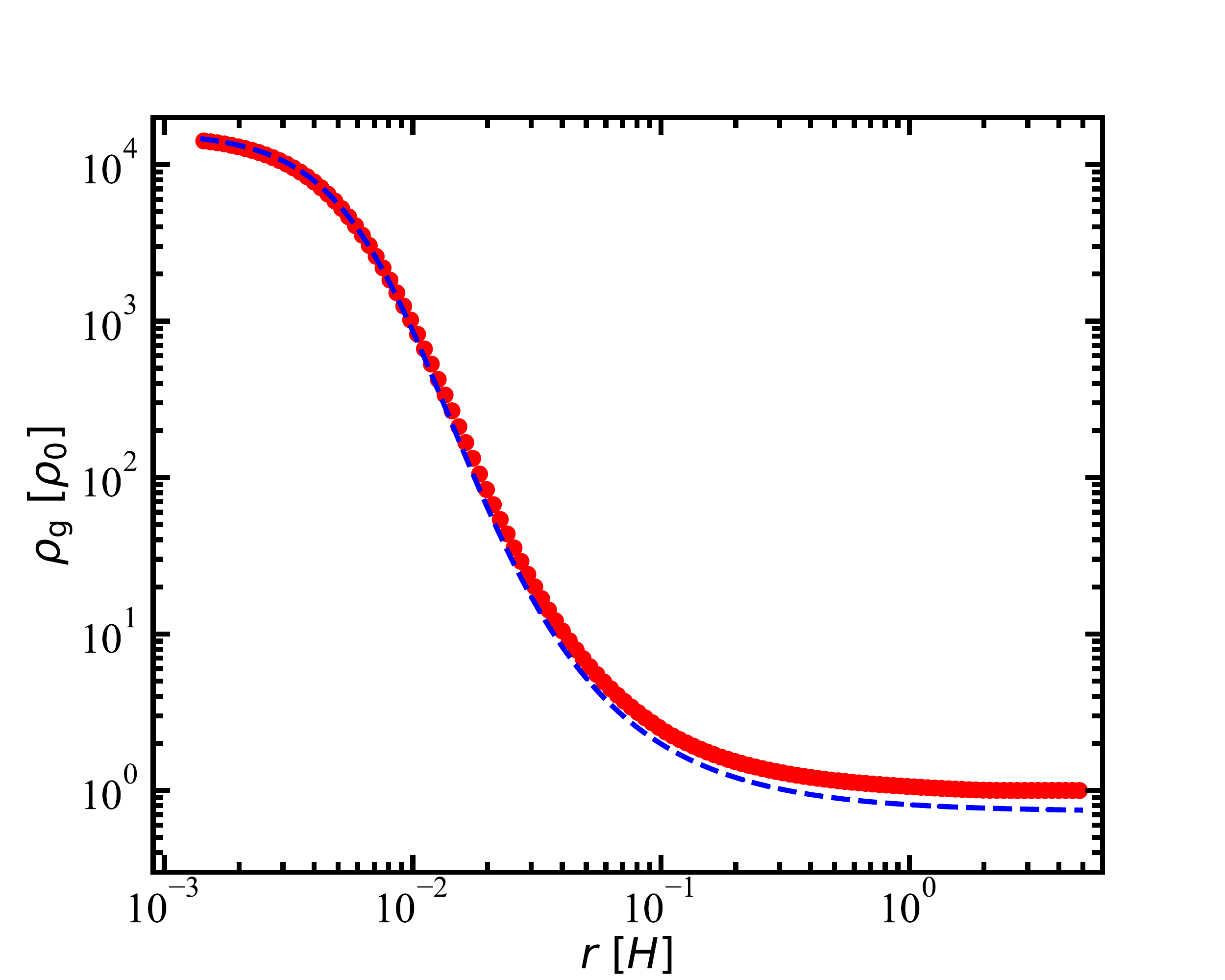}
	\caption{Gas density profile in the midplane ($z=0$) for $m=0.1$ is plotted as a function of distance from the center of the planet. The red dot shows the gas density averaged over the azimuthal direction ($\phi$) in the midplane for our numerical simulation. The blue dashed line corresponds to the analytical solution in Eq. (\ref{eq:denana}).}
	\label{fig:rhoanam01}
\end{figure}

To see the atmospheric density profile we plot the density profile averaged over the azimuthal direction ($\phi$) in the midplane (Fig. \ref{fig:rhoanam01}). 
The density at $r<0.1H$ significantly increases wi	th decreasing $r$, while the radial density slope at $r \lesssim 4\times 10^{-3}H$ becomes shallower due to the softening.
In a quasi-steady state, the density profile reaches the solution of the isothermal hydrostatic equilibrium in the one-dimensional analysis (see derivation in Appendix \ref{sec:appA}).
\rev{It should be noted that the density enhancement around the planet occurs even for $r \ga 0.1 H \approx 0.3 R_{\rm H} \approx R_{\rm B}$.}
\rev{Although the outer boundary of an atmosphere is conventionally set to the smaller of $R_{\rm H}$ and $R_{\rm B}$ \citep[e.g.,][]{inaba2003}, the atmospheric density enhancement occurs $r \la R_{\rm H}$ even if $R_{\rm H}>R_{\rm B}$}.

In the simulation, the density profile of the atmosphere is determined by the hydrostatic equilibrium in the cooling model that we use.
The profile is different from that for the growing planet, because the $\beta$ cooling model is too simple.
On the other hand, the cooling is effective in the vicinity of the planet.
Once the closed flow pattern forms the atmosphere, the flow outside the atmosphere would be almost independent of the cooling.
Taking into account our finding in the hydrodynamic simulations, we discuss the effect of the more realistic atmosphere in \S \ref{sec:realden}.

\subsection{Two-dimensional Orbital Calculations}

\subsubsection{Example of  2D Orbits}
\begin{figure*}[htb!]
	 \gridline{\fig{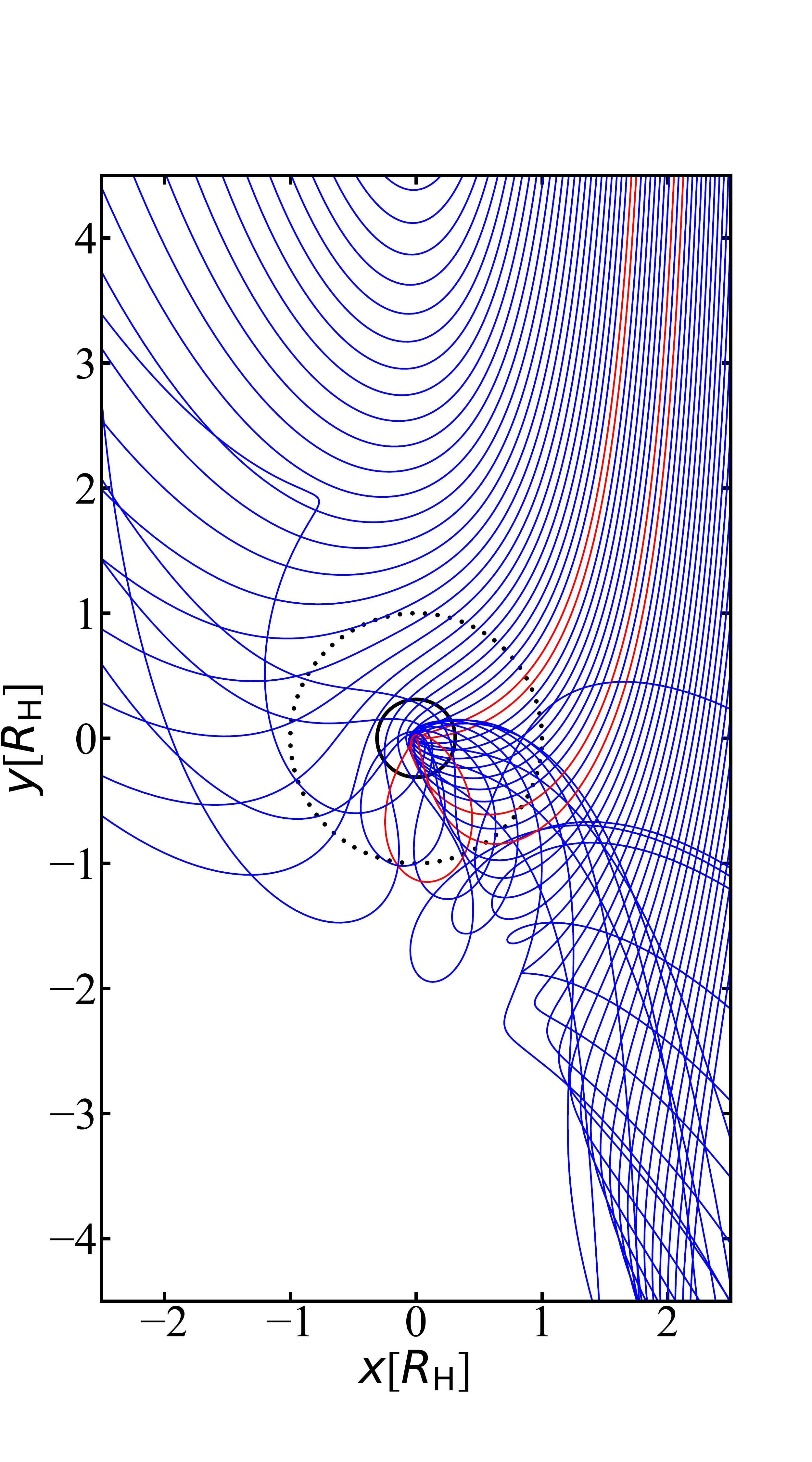}{0.3\textwidth}{(a)$St_0=1.0\times10^{13}$\label{fig:st1e13_orbit}}
	 \fig{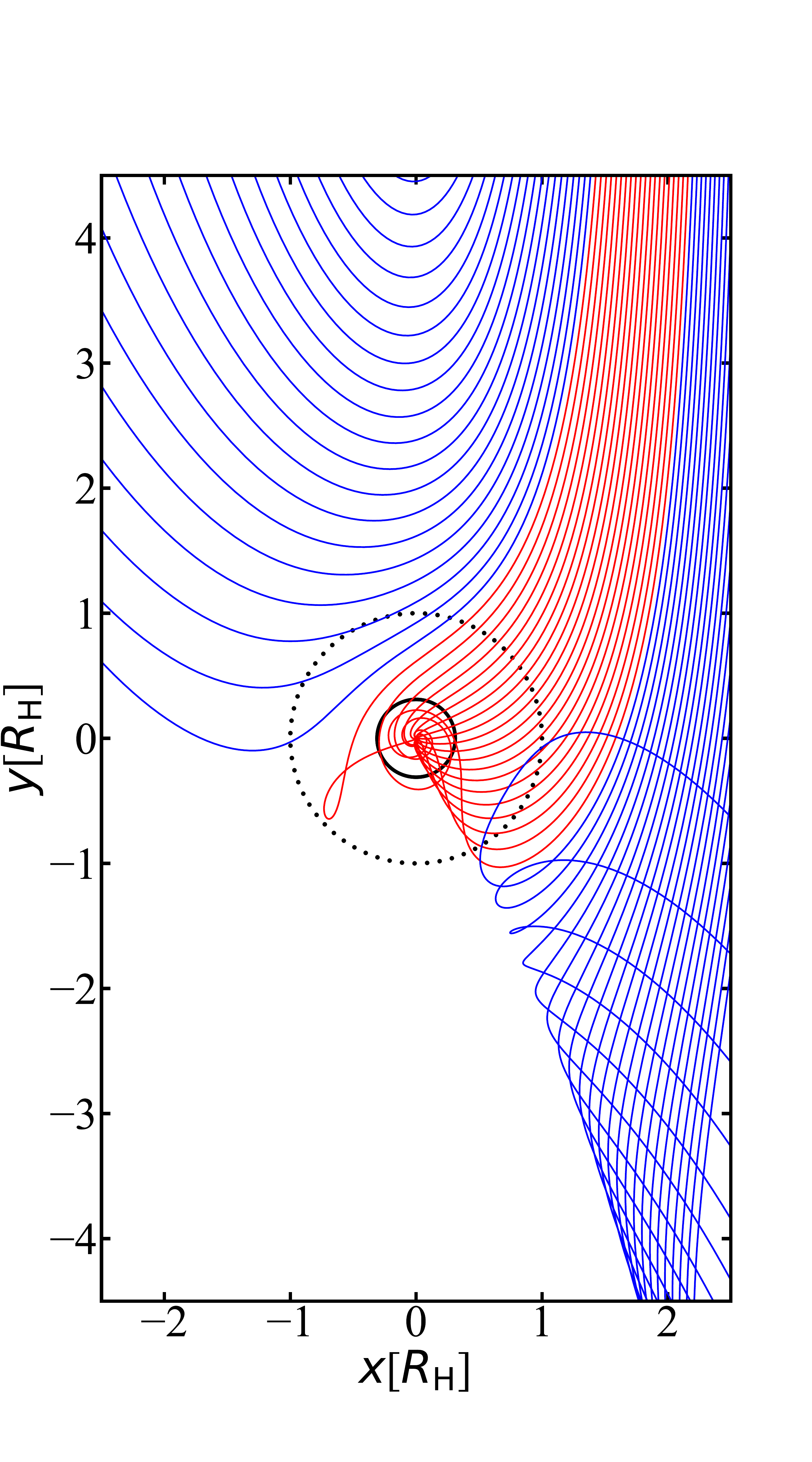}{0.3\textwidth}{(b)$St_0=1.0\times10^{2}$\label{fig:st1e2_orbit}}
	 }
 \gridline{\fig{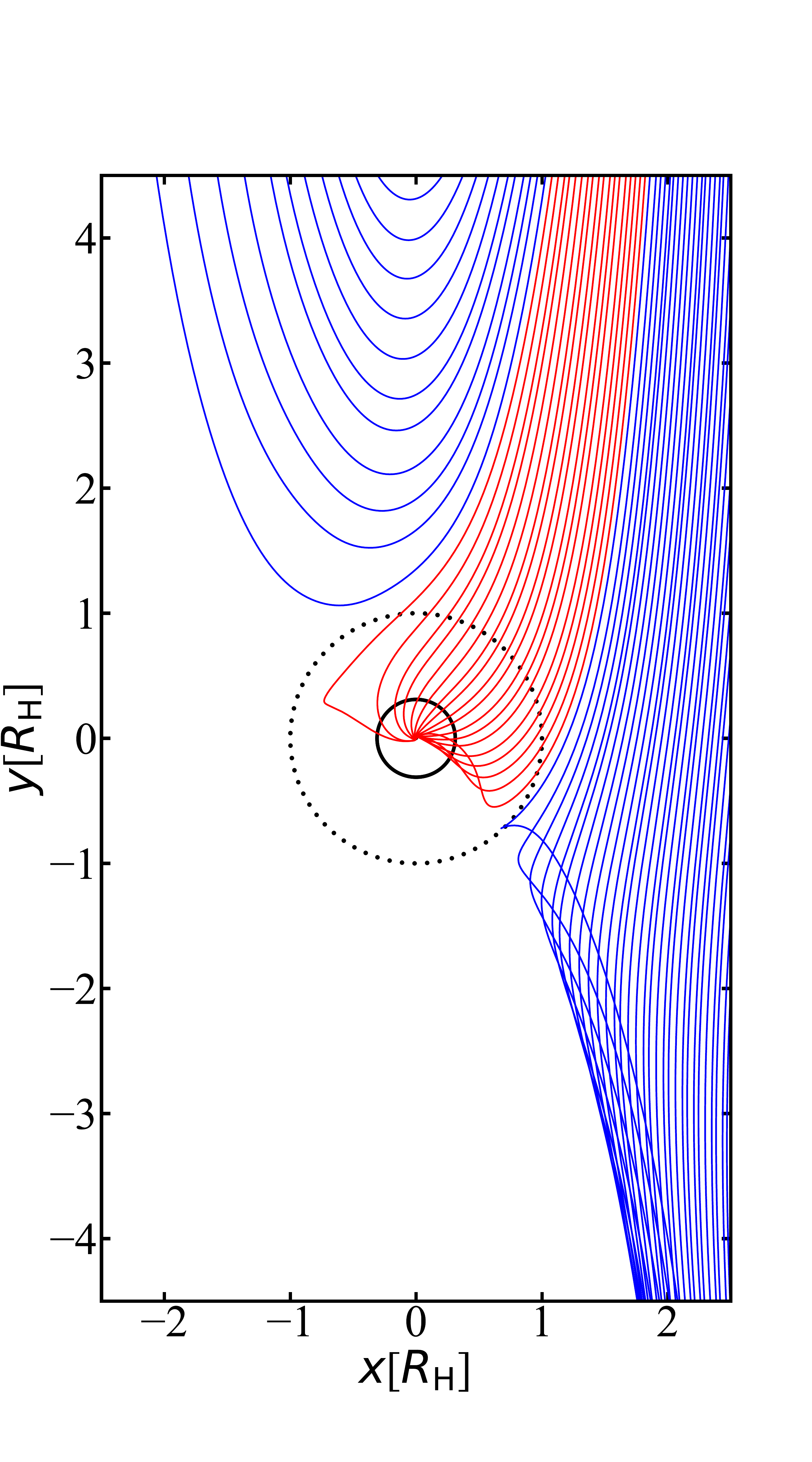}{0.3\textwidth}{(c)$St_0=1.0$\label{fig:st0_orbit}}
	 \fig{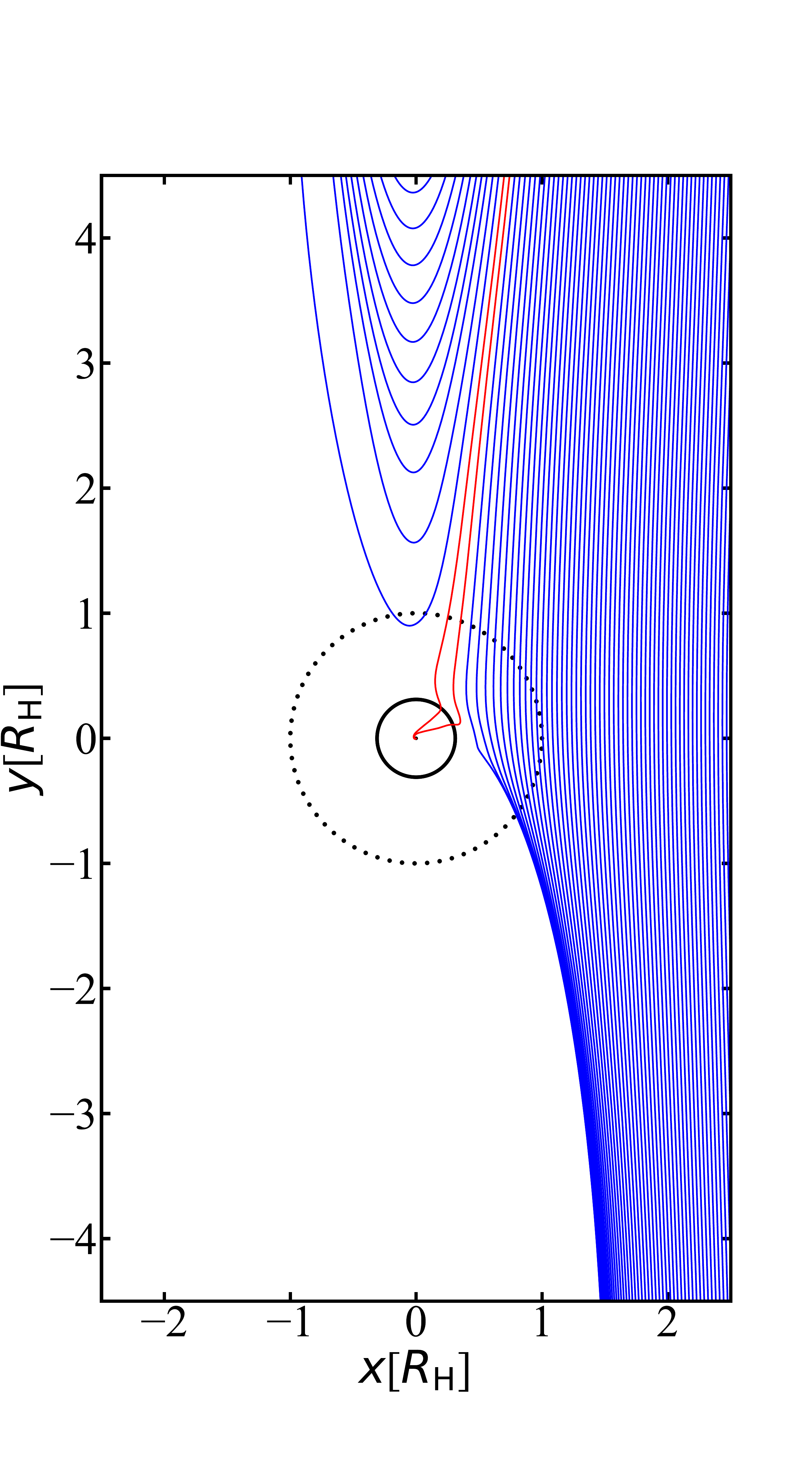}{0.3\textwidth}{(d)$St_0=1.0\times10^{-2}$\label{fig:st2_orbit}}
	 }
 \caption{Trajectories of particles with different Stokes parameter for $m=0.1$ are shown by solid lines. We restrict the motion of particles to 2D. The red and blue lines show the trajectories of particles that do and do not collide with the planet, respectively. The outer dotted  and inner black circles are the Hill sphere and the Bondi sphere, respectively. The interval of orbits at the starting points is $0.01H$.}\label{fig:orbit2d}
\end{figure*}

First of all, we perform two-dimensional orbital calculations in which particles have orbits in the midplane ($z=0$).
Figure \ref{fig:orbit2d} shows the trajectories of the particles with different $St_0$. 

For the large Stokes parameter, $St_0=1.0\times10^{13}$ (Fig. \ref{fig:orbit2d}a), the trajectories are almost the same as those in the gas-free case because particles are too large to be affected by the gas drag \citep{ida1989}.
Particles collide with the planet for the initial positions ($x_{\rm s}$) in three discrete bands (Fig. \ref{fig:orbit2d}a).

For $St_0 = 1.0 \times 10^2$ (Fig. \ref{fig:orbit2d}b), the orbits outside the Hill sphere is almost the same as those for $St_0= 1.0 \times 10^{13}$.
However, the particles feel strong gas drag in the atmosphere.
All the particles entering the Bondi sphere collide with the planet.

For $St_0 \gg 1$, the orbits of particles are almost independent of $St_0$ and similar to the gas-free ones unless $r<R_\mathrm{H}$.  
In the Hill sphere, the velocities of particles can be larger than the sound velocity and the gas density increases (see Fig. \ref{fig:rhoanam01}).
Therefore, gas drag effectively damps the kinetic energies of particles, which induces the capture of passing particles.

For $St_0=1.0$ (Fig. \ref{fig:orbit2d}c), the orbits of particles are similar to those for $St_0=1.0\times10^{2}$. \rev{The orbits of particles in $y<0$ are closer to the streamlines than those for $St_0=10^{2}$ are.}
\rev{In addition, all particles entering the Hill sphere accrete onto the planet.}

For $St_0=1.0\times10^{-2}$ (Fig. \ref{fig:orbit2d}d), the orbits are the almost same as the steady streamlines in Fig. \ref{fig:stream}.
The horseshoe width is much smaller than that for $St_0 \gtrsim 1$.
Particles are prevented from accreting onto the planet by the horseshoe flow and Keplerian shear flow, and the particles coming from the narrow band between the horseshoe and Keplerian shear flows are allowed to collide with the planet. This result is consistent with previous studies \citep{popovas2018, kuwahara2020a, kuwahara2020b, homma2020}.

\subsubsection{2D Collision Rate}
We calculate the two-dimensional specific collision rate, $P_\mathrm{col}=P_\mathrm{col,2D}$, based on the orbital calculations. The definition of $P_\mathrm{col,2D}$ is given by 
\begin{equation}
P_{\mathrm{col},2D}=2\int_0^{\infty}\Phi(x_\mathrm{s})|v_{0,y}|dx_\mathrm{s}=3\Omega \int_0^{\infty} \Phi(x_\mathrm{s}) x_\mathrm{s}dx_\mathrm{s},
\label{eq:pcol0}
\end{equation}
where $\Phi(x_\mathrm{s})=1$ if the particle collides with the planet and zero otherwise, and $v_{0,y}$ is $y$-component of the initial velocity in Eq. (\ref{eq:initialz}).
In order to account for the collision from both positive $x_{\rm s}$ and negative $x_{\rm s}$, we multiply the integral over positive $x_{\rm s}$ by two.

\begin{figure}[htb!]
	\centering
	\includegraphics[scale=0.3]{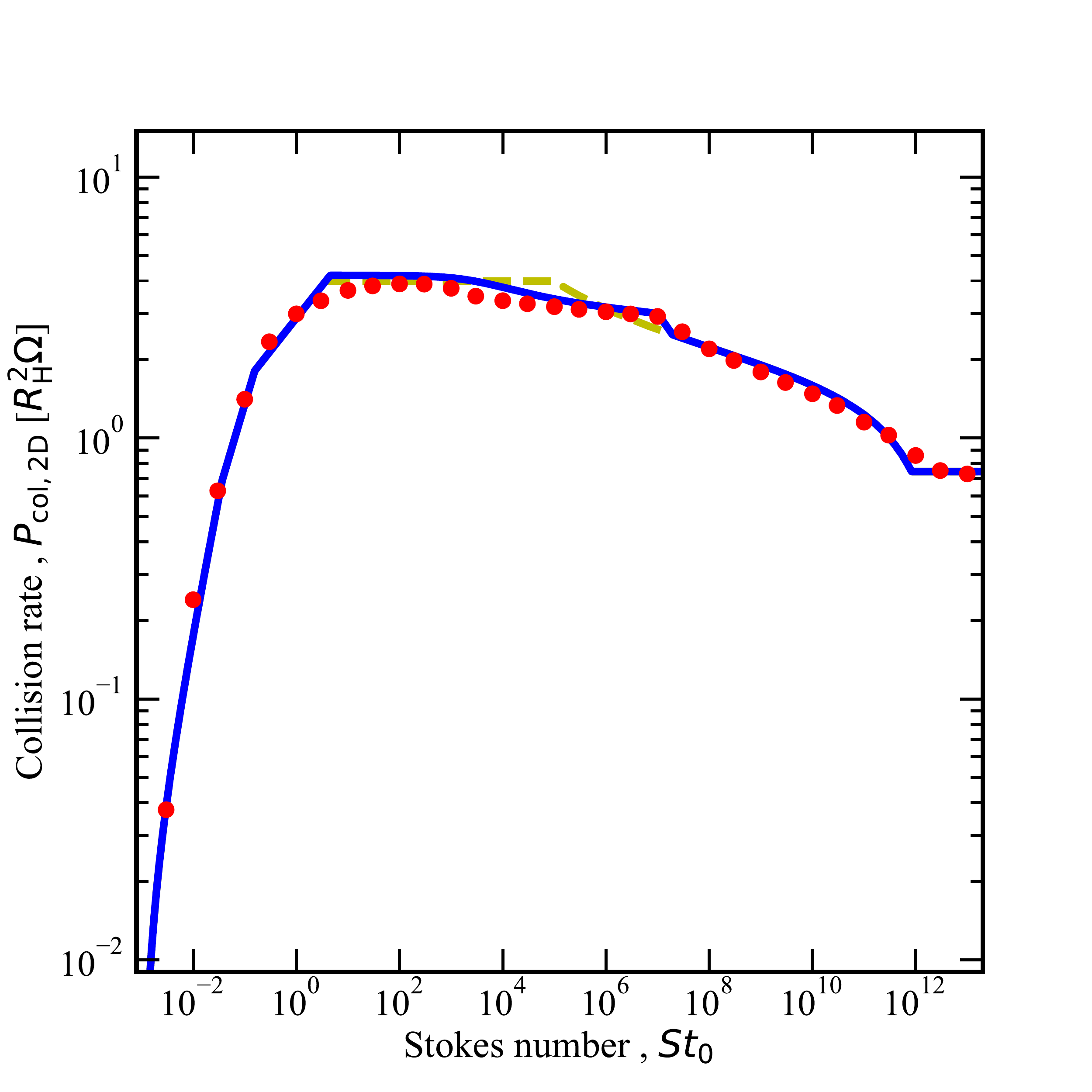}
	\caption{Two-dimensional specific collision rate, $P_\mathrm{col,2D}$, for $m=0.1$ as a function of $St_0$ given by Eq. (\ref{eq:stradii}). Red dots represent the collision rate obtained from the hydrodynamic and orbital calculations. The blue solid and yellow dashed lines indicate the detailed and simple analytical solutions given in Table \ref{tab:sumana}, respectively.}
	\label{fig:pcol2dm01}
\end{figure}

Figure \ref{fig:pcol2dm01} shows the collision rate  for $m=0.1$ as a function of $St_0$.
For $St_0 \gtrsim 10^{12}$, the collision rate converges to a constant value, which is the same as the gas-free limit \citep{ida1989, inaba2001}.

For $St_0 \gtrsim 1$, the collision rate increases with decreasing $St_0$.
For large $St_0$, particles are scattered away due to close encounters with the planet, so that $P_\mathrm{col,2D}$ is small.
\rev{However for small $St_0$, particles feel strong gas drag during close encounters because of high atmospheric density.}
\rev{Our hydrodynamic simulations indicate that density enhancement occurs even at $r \ga R_{\rm B}$ if $r < R_{\rm H}$ (see Fig. \ref{fig:rhoanam01}).}
\rev{For $ 1 \lesssim St_0 \lesssim 10^{2}$, particles can be captured at the outer edge of the atmosphere, $r \sim R_{\rm H}$.}
In addition, a close encounter with the planet accelerates the velocity, which can exceed the sound velocity. The super-sonic gas drag effectively reduces the kinetic energy prior to scattering.
Particles are then captured by atmosphere, which enhances $P_\mathrm{col,2D}$ \citep{inaba2003}.

For $St_0 \lesssim 1$, $P_\mathrm{col,2D}$ decreases with decreasing $St_0$.
Small particles are well coupled to the gas.
The flow pattern of the gas due to horseshoe and Keplerian shear limits the accretion of particles onto the planet (see Fig. \ref{fig:orbit2d}d).
Therefore, the collision rate sharply decreases with decreasing $St_0$.

\subsection{Three-dimensional Orbital Calculations}
\subsubsection{Example of 3D Orbits}
\begin{figure}[htb!]
	 \gridline{\fig{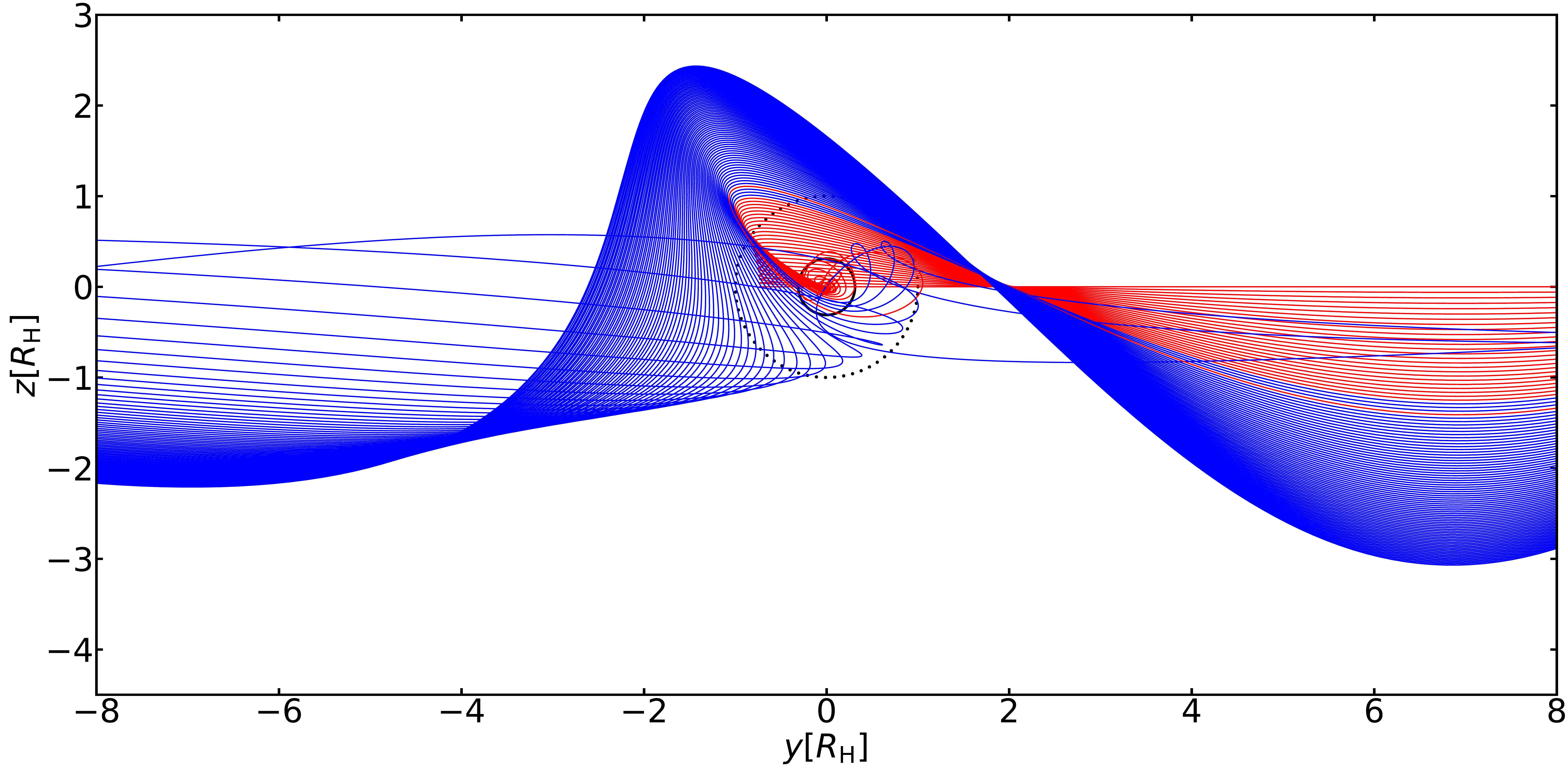}{0.45\textwidth}{(a)$St_0=1.0\times10^{3}$}\label{fig:3dorbit1e3}}
	 \gridline{\fig{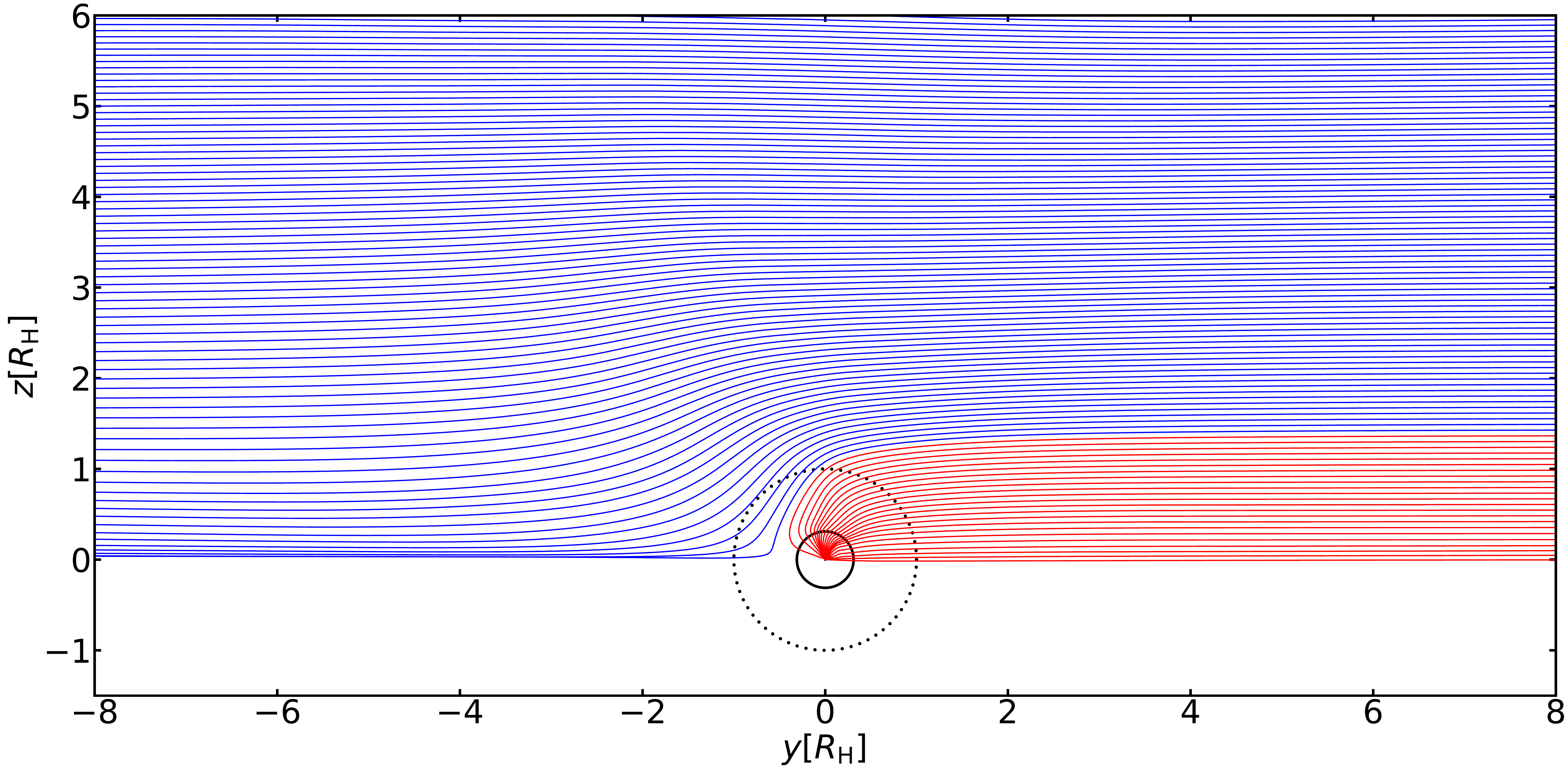}{0.45\textwidth}{(b)$St_0=1.0\times10^{-2}$}\label{fig:3dorbit2}
	 }
    \caption{Trajectories of particles for different $z_0$ with interval $0.02\ {\rm H}$ projected on the $y$-$z$ plane are shown by the solid lines for $St_0=1.0\times10^{3}$ with $x_{\rm s}=0.75\ H$ (a) and for $St_0 = 1.0\times10^{-2}$ with $x_{\rm s}=0.33\ H$ (b). The colors of solid lines and the outer dotted and inner circles are the same as Fig. \ref{fig:orbit2d}. Note that the ranges of $z$ are different in Pannels (a) and (b).}\label{fig:3dorbityz}
\end{figure}

First, we show the case of large particles (for $St_0 > 1$). For $i a_{\rm s}$ or $z_0$ $\ll R_{\rm H}$, the 3D orbits projected on the $x$-$y$ plane are almost the same as the 2D orbits. 
Figure \ref{fig:3dorbityz}a shows the orbits of particles projected on the $y$-$z$ plane for $St_0=1.0\times10^3$.
For $i\gg R_{\rm H}/a_{\rm s}$, particles pass through the planet.
Particles passing near the planet can be accreted onto the planet.

Figure \ref{fig:3dorbityz}b shows the orbits of the smaller particles projected on the $y$-$z$ plane for $St_0=1.0\times10^{-2}$.
Particles can be accreted if they pass near the planet as well as the case of $St_0 \gtrsim 1$.

\subsubsection{3D Collision Rate}
We calculate the three-dimensional specific collision rate.
We then need the vertical distribution of particles for the calculation.

For $St_0 > 1$, we set the Rayliegh type distribution for inclinations \citep{idamakino1992a}.
Assuming the random orbital phases, we obtain the collision rate as, 
\begin{eqnarray}
P_{\rm col,3D}(i^*)=\frac{6 \Omega}{\pi i^{*2}} \int _0^{\infty} \int_0^{\pi} \int_0^{z_{0,\mathrm{max}}/a_{\rm s}} \Phi(x_{\rm s}, i , \omega) x_{\rm s} i  \nonumber
\\
\times \mathrm{exp} \left(-\frac{i^2}{i^{*2}}\right) dx_{\rm s} d\omega di \nonumber \\,
\end{eqnarray}
where $i^{*}$ is the dispersion of $i$ and $z_{0,\mathrm{max}}$ is the upper limit of $z_{0}$.
Note that the vertical distribution of particles under this assumption corresponds to a Gaussian distribution for $z_0$ with the scale hight of  $a_{\rm s} i^*/\sqrt{2}$ (see Appendix \ref{sec:appB}).

For $St_0<1$, we assume the Gaussian distribution function of $z_0$.
The collision rate is given by 
\begin{eqnarray}
P_{\rm col,3D}(H_{\rm d}) = \frac{3 \Omega}{\sqrt{2 \pi} H_{\rm d}}\int_0^{\infty} \int_0^{\infty} \Phi(x_{\rm s},z_0) \nonumber \\
\times x_{\rm s} \mathrm{exp}\left(-\frac{z_0^2}{2 H_{\rm d}^{2}}\right) dx_{\rm s} dz_0,
\end{eqnarray} 
where $H_{\rm d}$ is the scale hight of particles.

\begin{figure}[htb!]
\centering
	\includegraphics[scale=0.28]{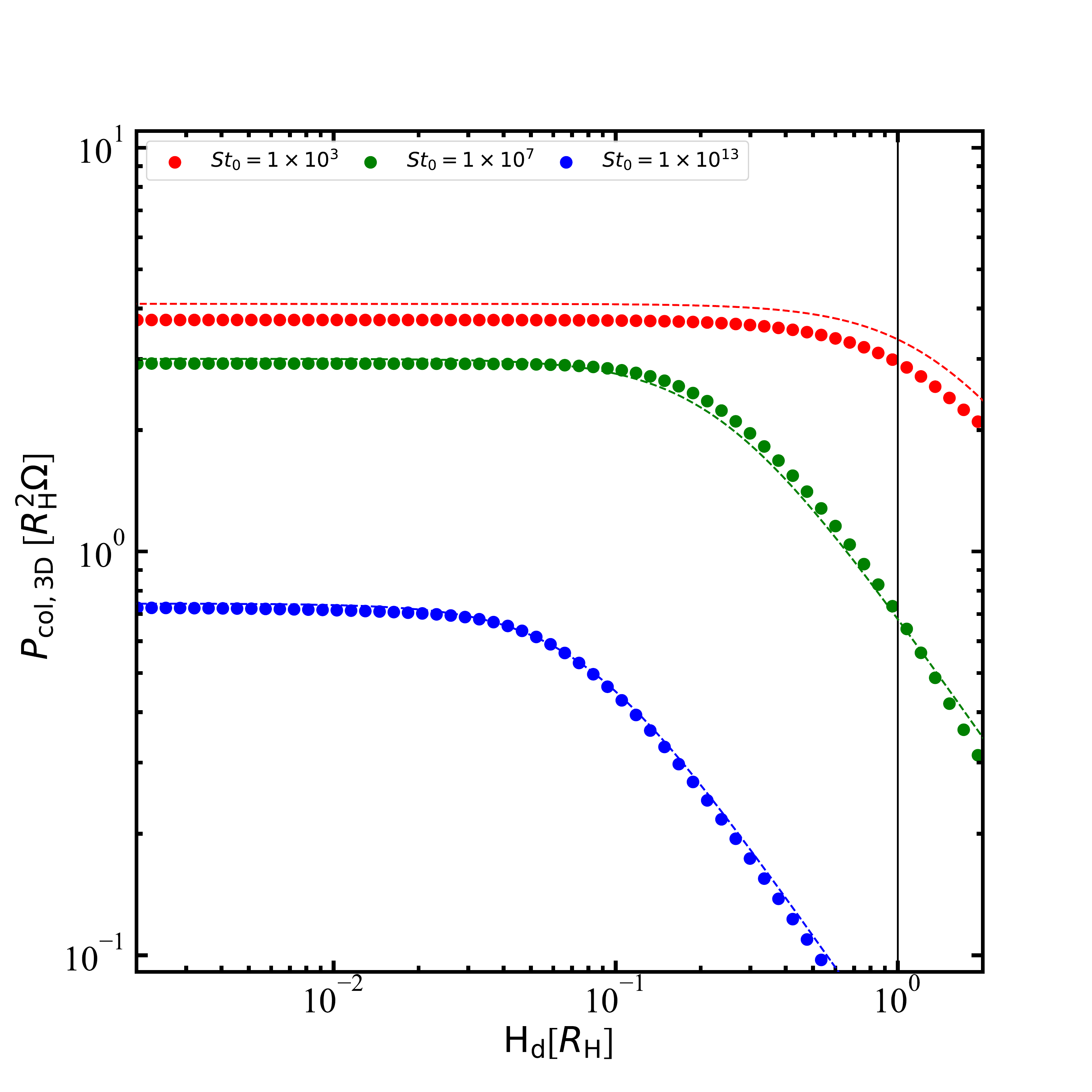}
	\caption{Three-dimensional specific collision rates, $P_{\rm col,3D}$, for $m=0.1$ as a function of the particle scale hight $H_{\rm d} =  a i^{*}/\sqrt{2}$, where $i^{*}$ is the dispersion of inclinations and $a$ is the semimajor axis of the planet. Dots represent the collision rates obtained via the hydrodynamic and orbital simulations  for $St_0=10^{3}$(red), $10^{7}$(green), $10^{13}$(blue). Dashed lines indicate the analytical solutions given in Table \ref{tab:sumana}. The vertical solid black line is the Hill radius for reference.}
	\label{fig:pcol3dlargem01}
\end{figure}

Figure \ref{fig:pcol3dlargem01} shows $P_{\rm col,3D}(i^*)$ for $St_0>1.0$ as a function of $H_\mathrm{d}=a_{\rm s} i^{*}/\sqrt{2}$, using the relation between $H_{\rm d}$ and $i^{*}$ given in Eq. (\ref{eq:hd_iasta}).
For $H_{\rm d}\ll R_\mathrm{H}$, $P_{\rm col,3D}$ is independent of $H_{\rm d}$ and almost the same as $P_{\rm col,2D}$. For $H_{\rm d}\gg R_\mathrm{H}$, $P_{\rm col,3D}$ decreases with $H_{\rm d}$.
This is caused by the passing by without collisions for high inclination particles (see Fig. \ref{fig:3dorbityz}a).

\begin{figure}[htb!]
	\centering
	\includegraphics[scale=0.28]{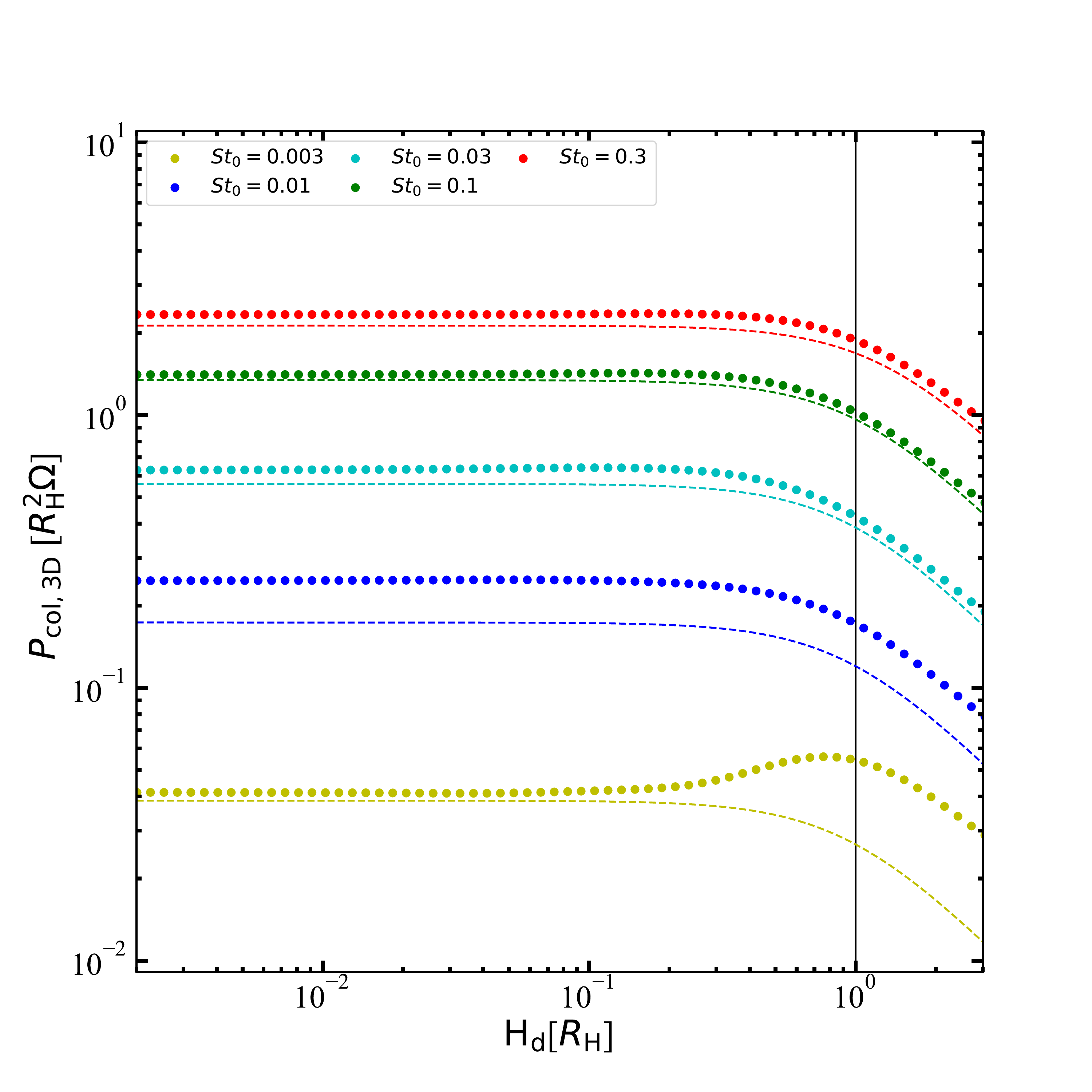}
	\caption{Same as Fig. \ref{fig:pcol3dlargem01} but for $St_0 < 1$.}
	\label{fig:pcol3dsmallm01}
\end{figure}

Figure \ref{fig:pcol3dsmallm01} shows  $P_{\rm col,3D}(H_{\rm d})$ for $St_0<1.0$.
The dependence of $P_{\rm col,3D}$ on $H_{\rm d}$ is similar to that for $St_0>1$.
For $St_0=0.003$, $P_{\rm col,3D}(H_{\rm d})$ increases around the $H_{\rm d}=R_{\rm H}$.
This increase is caused by the vertical gas flow that enters from high latitudes as seen in Fig. \ref{fig:vz} \citep[see also][]{kuwahara2020a}.

\section{Analytic Formulae for $P_{\rm col}$}\label{sec:resultana}

\subsection{Analytic Formula of the 2D Collision Rate}
In this subsection, we derive the analytic formulae for the two-dimensional collision rate.
\rev{The dominant effects determining the accretion rate depend on $St_0$. Therefore, we give different assumptions depending on $St_0$ to derive analytic formulae below. 
In \S.~\ref{sc:gas_free}, 
we introduce the previous study for the analytic formula without gas drag. 
In \S.~\ref{sec:pcol2datm}, we consider the atmospheric capture regime, where particles with $St_0 \ga 1$ entering atmospheres are captured due to the energy losses by gas drag in atmospheres. In \S.~\ref{sc:settling}, 
we consider particles with $St_0 \la 1$
settle down to a planet with terminal velocities during a close encounter with the planet. 
In \S.~\ref{sec:hoeffect}, we consider particles with $St_0 \ll 1$ are strongly affected by the horseshoe and shear flows of gas. 
}

\subsubsection{Gas Free Limit}
\label{sc:gas_free}

\rev{First, we confirm the analytic formula in the gas free limit that means particles and $St_0$ are extremely large.}
The two-dimensional collision rate for the gas-free limit was derived in the previous studies, given by \citep{ida1989, inaba2001}
\begin{equation}
\frac{P_{\mathrm{col},\mathrm{free}0}(R_{\rm pl})}{R_\mathrm{H}^2 \Omega}=11.3\sqrt{R_\mathrm{pl}/R_\mathrm{H}},
\label{eq:pcolfree}
\end{equation}
where $R_\mathrm{pl}$ is the radius of the planet.
This formula is in agreement with our simulations for the huge Stokes number.
\rev{Note that Eq. (\ref{eq:pcolfree}) is valid for $R_{\rm pl} \ll R_{\rm H}$.}
\rev{We improve the formula for $R_{\rm pl} \sim R_{\rm H}$ in \S \ref{sec:pcol2datm}}.

\subsubsection{Atmospheric Capture}\label{sec:pcol2datm}
\rev{Second, we derive the analytic formula for the atmospheric capture regime.}
Particles entering planetary atmosphere have high velocities due to planetary gravity.
We estimate the terminal velocity of particles at $r=R_\mathrm{B}$ as
\begin{eqnarray}
u &\sim& \frac{GM_\mathrm{p}}{R_{\rm B}^2}\frac{St_0}{\Omega} \nonumber \\
&=& \frac{St_0}{m}c_\mathrm{s}.
\label{eq:velterm}
\end{eqnarray}
For $St_0\ga m$, the velocity can be greater than the sound velocity, so that the super-sonic gas drag given by the second term in Eq. (\ref{eq:cd}) is effective in the atmosphere.

The kinetic energies of particles are damped by gas drag in the atmosphere.
Once the particles' energies are smaller than the gravitational energy at the Hill radius, the orbits of the particles are bound by the planet.
The kinetic energy at infinity is negligible.
The condition for capture of particles is then given by 
 \begin{equation}
\Delta E<-\frac{GM_\mathrm{p}m_\mathrm{p}}{R_\mathrm{H}},
\label{eq:inabaikoma}
\end{equation}
where $\Delta E$ is the energy loss due to gas drag in the atmosphere.
We estimate $\Delta E$ from the integral of the energy loss along the particle orbit in the atmosphere with radius $R_{\rm atm}$, given by
\begin{eqnarray}
\Delta E(q)  &\approx& - \zeta  \int \pi r_\mathrm{p}^2 \rho_\mathrm{atm}(r) u^3 \ dt \nonumber \\
&=& - 4 \zeta \pi G M_\mathrm{p} r_\mathrm{p}^2 \int^{R_\mathrm{atm}}_{q}\frac{\rho_\mathrm{atm}(r)}{\sqrt{r(r-q)}}\ dr, \nonumber \\
\label{eq:dele}
\end{eqnarray}
\rev{where $\zeta$ is the number of close encounters between the planet and the particle and we assume a parabolic orbit with pericenter distance $q$ for the particle and then we use the relations $u^{2}=2 G M_{\rm p}/r$ and $dr/dt= \sqrt{2 G M_{\rm p} (r-q)}/r$ for the derivation.}
We multiply two because the integral range is the half of a parabola.
The number of close encounters and the atmospheric radius are determined according to our simulation and we adopt $\zeta=2$ and $R_{\rm atm}=0.5\ R_{\rm H}$.
The following condition gives the captured radius $R_\mathrm{cap}$
\begin{equation}
\Delta E(R_\mathrm{cap})=-\frac{G M_\mathrm{p} m_\mathrm{p}}{R_\mathrm{H}}.
\label{eq:rcap}
\end{equation}
We derive $R_\mathrm{cap}$ satisfing Eq. (\ref{eq:rcap}).

\begin{figure}[htb!]
	\centering
	\includegraphics[scale=0.25]{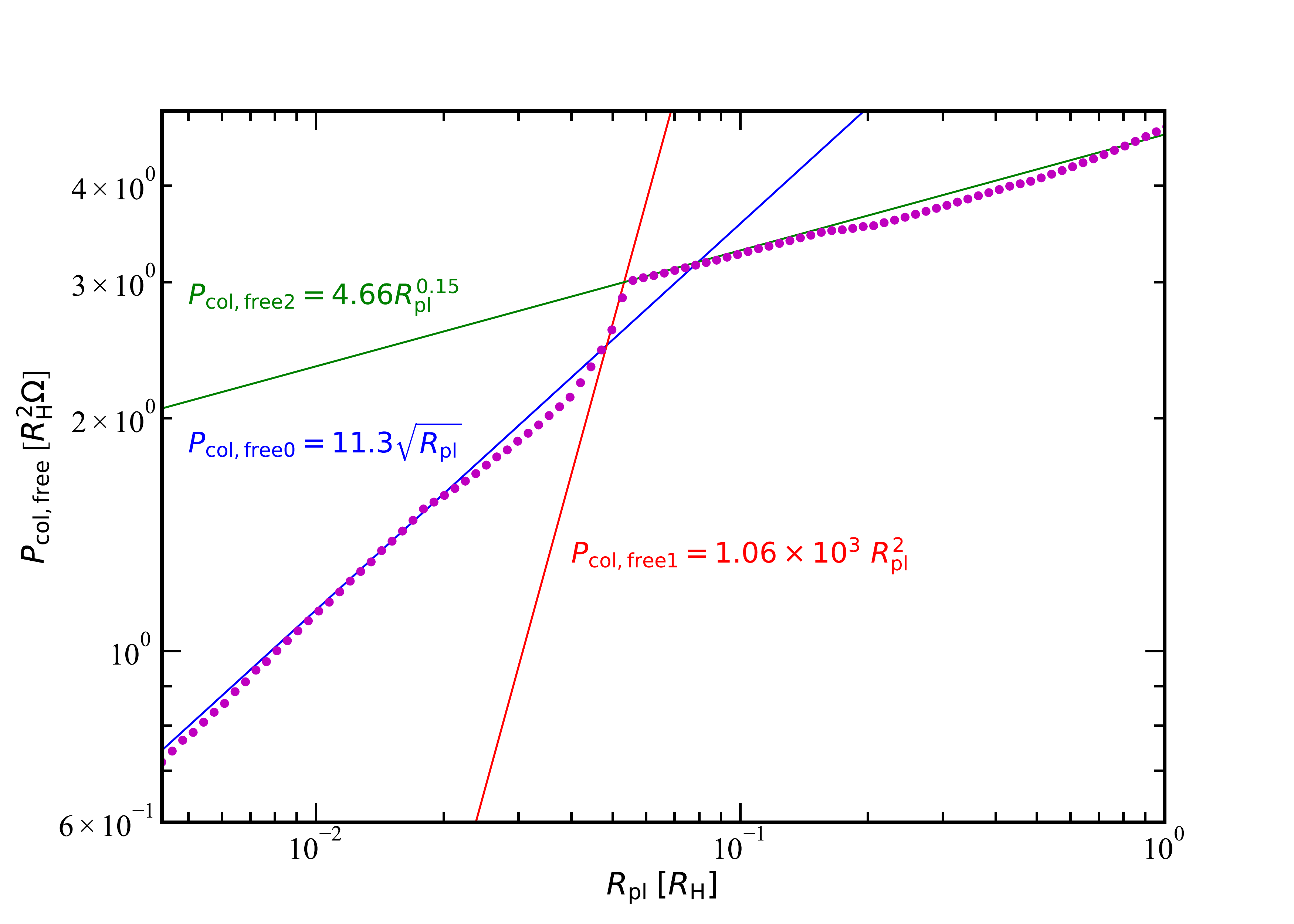}
	\caption{$P_{\rm col,2D}$ is shown as a function of the planetary radius in the gas-free case.}
	\label{fig:rmin_pcol}
\end{figure}

\rev{According to \citet{inaba2003}, we obtain $P_{\rm col}$ from $P_{\rm col,free0}$ using $R_{\rm cap}$ instead of $R_{\rm pl}$.}
However, $P_{\rm col,free0}$ is valid for $R_{\rm pl} \ll R_{\rm H}$.
Figure \ref{fig:rmin_pcol} shows the collision rate in the gas-free case as a function of the planetary radius. For $R_{\rm pl} > 0.05R_{\rm H}$, we obviously need to modify the formula.
Therefore, we give the collision rate
\begin{equation}
P_{\mathrm{col},\mathrm{free}}= \mathrm{MIN}\left[\mathrm{MAX}\left(P_{\rm col,free0}, \ P_{\rm col,free1} \right) ,\ P_{\rm col,free2} \right],
\label{eq:pcolfree2}
\end{equation}
where
\begin{eqnarray}
\frac{P_{\rm col,free1}(R_{\rm pl})}{R_{\rm H}^2\Omega}&=&1.06\times10^{3}(R_{\rm pl}/R_{\rm H})^{2},\\
\label{eq:pcolf1}
\frac{P_{\rm col,free2}(R_{\rm pl})}{R_{\rm H}^2\Omega} &=& 4.66 (R_{\rm pl}/R_{\rm H})^{0.15}.
\label{eq:pcolf2}
\end{eqnarray}
For the atmospheric collision rate, we use $R_{\rm cap}$ instead of $R_{\rm pl}$ and then obtain
\begin{equation}
P_{\rm col,atm} = P_{\rm col,free}(R_{\rm cap}).
\label{eq:pcolatm}
\end{equation}

\subsubsection{Settling}
\label{sc:settling}

\rev{Third, we derive the analytic formula for the settling regime.}
\rev{There are two regimes for settling; the cases of the supersonic gas drag and the Stokes gas drag.}

For $St_0 \lesssim 1$, \citet{ormel2010} derived $P_{\rm col}$ for constant $St$ in the Keplerian shear flow.
They obtained the analytic formula by comparing the encounter time with the settling time.
For a particle with an impact parameter $x_{\rm s}$, the encounter time is estimated to be $t_\mathrm{enc}=x_{\rm s}/|v_{0,y}|$, where $|v_{0,y}|=3\Omega x_{\rm s}/2$ is the encounter velocity.
The settling time needed for particles to \rev{settle down} to the planet is given by $t_\mathrm{set}=x_{\rm s}/u_\mathrm{term}$, where $u_\mathrm{term}$ is the terminal velocity determined by the force balance between the gravitational force and gas drag.

\rev{As explained above, for $St_0\ga m$, the velocity of the particle approaching the planet exceeds the sound velocity}, so that the supersonic gas drag, the second term in Eq. (\ref{eq:cd}), is effective. The equation of force balance between the gravity and gas drag is given by
\begin{equation}
\frac{GM_\mathrm{p}m_\mathrm{p}}{r^2}=\pi r_\mathrm{p}^2 \rho_\mathrm{g} u_\mathrm{term}^2,
\end{equation}
so that the terminal velocity is expressed by
\begin{equation}
u_\mathrm{term}=\sqrt{\frac{4G M_\mathrm{p}\rho_\mathrm{p}}{3 \rho_\mathrm{g} x_{\rm s}^2}}\left(\frac{9\rho_\mathrm{g}c_\mathrm{s}l_\mathrm{mfp}}{4 \rho_\mathrm{p}}\frac{St_0}{\Omega}\right)^{1/4}.
\end{equation}

For $t_{\rm enc} \gtrsim t_{\rm set}$, particles are accreted onto the planets.
The accretion condition is given as
\begin{equation}
C_1 \frac{x_{\rm s}}{|v_{0,y}|}>\frac{x_{\rm s}}{u_\mathrm{term}},
\end{equation}
where $C_1$ is the constant value on the order of unity.
The impact parameter $x_{\rm s}$ required for the settling is given by
\begin{equation}
x_{\rm s} < x_{\rm ss} = \left(\frac{8}{3}\right)^{1/4} \sqrt{C_1} \left(\frac{\rho_\mathrm{p}}{\rho_\mathrm{g}}\frac{c_\mathrm{s}}{R_\mathrm{H}\Omega}\frac{l_\mathrm{mfp}}{R_\mathrm{H}}St_0\right)^{1/8} R_{\rm H}.
\label{eq:xss}
\end{equation}
If $x_{\rm ss}$ is much larger than the half width of the horseshoe orbit, $P_{\rm col}$ is given by
\begin{equation}
P_{\rm col}=3 \Omega x_{\rm ss}^2.
\label{eq:sspcol}
\end{equation}
Using Eqs. (\ref{eq:xss}) and (\ref{eq:sspcol}), we obtain
\begin{equation}
\frac{P_{\mathrm{col},\mathrm{ss}}}{R_\mathrm{H}^2\Omega} =2\sqrt{6} C_1\left(\frac{\rho_\mathrm{p}}{\rho_\mathrm{g}}\frac{c_\mathrm{s}}{R_\mathrm{H}\Omega}\frac{l_\mathrm{mfp}}{R_\mathrm{H}}St_0\right)^{1/4}.
\label{eq:pcolss}   
\end{equation}

\par
\rev{On the other hand, if the terminal velocity of particles is smaller than the sound speed, we consider $St \approx St_0$ even in the encounter.}
\rev{In that case, the Stokes gas drag, the first term in Eq. (\ref{eq:cd}), is effective.}
The terminal velocity is given by
\begin{equation}
u_\mathrm{term}=\frac{GM_\mathrm{p}}{r^2}\frac{St_0}{\Omega}.
\label{eq:settlingvel}
\end{equation}
Same as above, Eq. (\ref{eq:settlingvel}) gives
\begin{equation}
x_{\rm ss} = (2C_1 St_0)^{1/3} R_\mathrm{H}.
\label{eq:xset}
\end{equation}
Using Eqs. (\ref{eq:sspcol}) and (\ref{eq:xset}), we obtain the collision rate,
\begin{equation}
\frac{P_{\mathrm{col},\mathrm{set}}}{R_\mathrm{H}^2\Omega}=3\left(2C_1 St_0\right)^{2/3}.
\label{eq:pcolset}
\end{equation}
According to the our simulations, $C_1=1.5$.


\subsubsection{Effects of the Horseshoe and Outflow around the Bondi Radius}\label{sec:hoeffect}
\rev{Finally, we derive the analytic formula considering the effects of the horseshoe and outflow around the Bondi radius.}
In our simulation, the outflow with a few percent of the sound speed is found around the Bondi radius, as shown in the previous study \citep{kuwahara2019}.
For smaller $St_0$, the horseshoe and outflow around the Bondi radius is effective. 
The collision rate is then given by $P_\mathrm{col} = 2 \Delta x_{\rm s} |v_{0,y}|$, where $\Delta x_{\rm s}=|x_{\rm ss}-r_\mathrm{HS}|$ is the difference between the impact parameter and the half horseshoe width $r_\mathrm{HS}$.
The particles passing around $r \sim R_{\rm B}$ are accreted onto the planet if they can enter the Bondi radius \rev{(see Fig. \ref{fig:orbit2d}d).}
We consider the passing particles are distributed from $r=R_{\rm B}$ to $r=R_{\rm B}+ \Delta r$.
The mass conservation gives $\Delta x_{\rm s} |v_{0,y}|=\Delta r v_{\theta}$.
If the particle at $r=R_{\rm B}+\Delta r$ enters the Bondi radius with the radial drift of $\Delta r$ during the encounter, the particle then accrete onto the planet.
Therefore, the accretion condition is given by the passing time comparable to or longer than the drift time; $R_{\rm B}/v_{\theta} \gtrsim \Delta r / v_r$.
We thus estimate $P_{\rm col}$ as 
\begin{equation}
P_{\rm col} = 2R_{\rm B}v_r.
\label{eq:hopcol}
\end{equation} 
Assuming the outflow velocity of $\xi c_{\rm s}$ with $\xi \ll 1$, $v_r$ is expressed by the terminal velocity $u_{\rm term}$ and the outflow velocity $\xi c_{\rm s}$, given by
\begin{equation}
v_r=\frac{GM_{\rm P}}{r^2}\frac{St_0}{\Omega}-\xi c_{\rm s}.
\label{eq:hovel}
\end{equation}
Using Eqs. (\ref{eq:hopcol}) and (\ref{eq:hovel}), we obtain
\begin{equation}
\frac{P_{\mathrm{col},\mathrm{ho}}}{R_\mathrm{H}^2 \Omega} = 2\frac{R_\mathrm{B}}{R_\mathrm{H}}\left[\frac{3 St_0}{\left(R_\mathrm{B}/R_\mathrm{H}\right)^2}-\xi \sqrt{\frac{3}{\left(R_\mathrm{B}/R_\mathrm{H}\right)}}\right].
\label{eq:pcolho}
\end{equation}
\rev{We adopt $\xi = 0.1m$, based on the results of hydrodynamic simulations for $m=0.03-0.1$.}
\rev{Note that the radial outflow velocity determined by $\xi$ is much smaller than the flow velocity around $r \sim R_{\rm B}$ derived by \citet{kuwahara2019}. We use the $m$ dependence of $\xi$ given by their formula. 

}

\subsection{Analytic Formulae of 3D Collision Rates}\label{sec:pcol3d}
In this subsection, we focus on the three-dimensional specific collision rate.

For $St_0 > 1$, the $i^*$ dependence of $P_{\rm col}$ is given in \citet{inaba2001}.
Here, we consider the capture radius instead of the planetary radius according to \S \ref{sec:pcol2datm} \citep[e.g.,][]{inaba2003}. We then obtain
\begin{eqnarray}
\frac{P_\mathrm{col,3D}(i^*)}{R_\mathrm{H}^2\Omega} &=&\left\{\left(\frac{P_{\rm col,2D}}{R_\mathrm{H}^2\Omega}\right)^{-2} \right. \nonumber \\
&&\left. +\left[\frac{R_{\rm cap}^2}{4\pi a i^* R_\mathrm{H}}\left(17.3+\frac{232 R_\mathrm{H}}{R_{\rm cap}}\right)\right]^{-2}\right\}^{-1/2},
\label{eq:3danalarge}
\end{eqnarray}
where $R_{\rm cap}$ is the captured radius obtained from Eq. (\ref{eq:rcap}).

For $ St_0 < 1$, $P_{\rm col , 3D} \approx P_{\rm col , 2D}$ for $H_{\rm d} \ll R_{\rm H}$, while $P_{\rm col , 3D} \propto x_{\rm ss}/H_{\rm d}$ for $H_{\rm d} \gg R_{\rm H}$ (see Fig. \ref{fig:pcol3dsmallm01}). 
\rev{Therefore, we empirically give}
\begin{equation}
P_\mathrm{col,3D}(H_{\rm d}) = \left[\left( P_{\mathrm{col,2D}}\right)^{-2}+\left(P_{\mathrm{col,2D}}\frac{x_{\rm ss}}{0.65H_\mathrm{d}}\right)^{-2}\right]^{-1/2},
\label{eq:3danasmall}
\end{equation}
where $P_{\rm col,2D}$ is given by the smallest of Eqs. (\ref{eq:pcolss}), (\ref{eq:pcolset}), and (\ref{eq:pcolho}) and $x_{\rm ss}$ is determined accordingly (see Table \ref{tab:sumana}).
If $P_{\rm col,2D}$ is given by Eq. (\ref{eq:pcolho}),  we set $x_{\rm ss}=2R_{\rm B}$.

\rev{Figures \ref{fig:pcol3dlargem01}, \ref{fig:pcol3dsmallm01}, \ref{fig:m005}, and \ref{fig:m003} show the analytical solution and simulation results. }
These formulae are in agreement with our simulation for almost all Stokes numbers, \rev{but for $St_0=0.003$ in $m=0.1$} the mean collision rate is underestimated for $H_{\rm d} \gtrsim R_{\rm H}$.
This is the result of vertical gas flow that enters from high latitudes \citep{kuwahara2020a, homma2020}.
This flow brings the particle that accretes onto the planet, so that the collision rate is maintained or increased.

\subsection{Summary of Equations for $P_{\rm col}$} \label{sec:summarypcol}
\begin{figure*}[htb!]
	 \gridline{
	 \fig{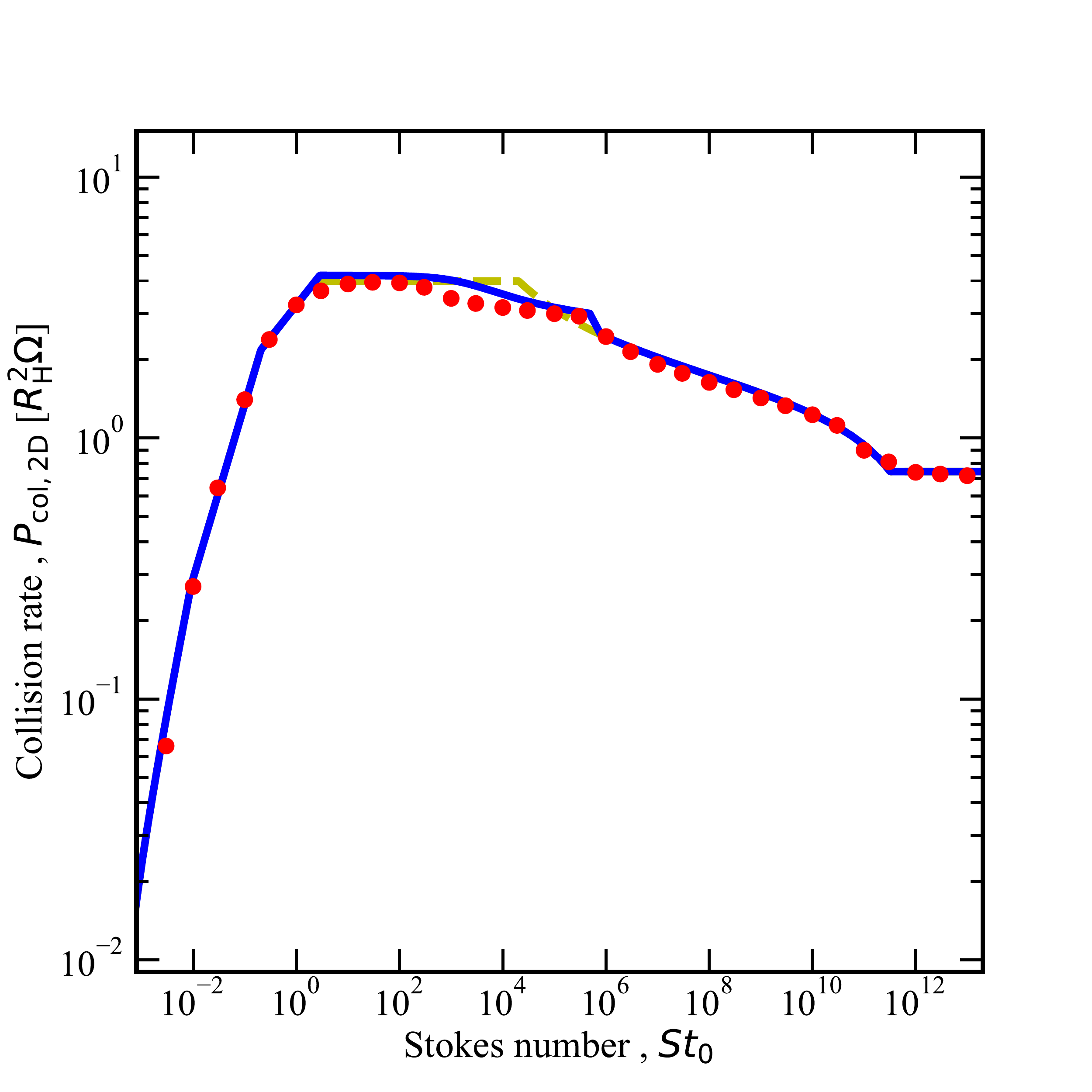}{0.33\textwidth}{(a) $P_{\rm col,2D}$}
	 \fig{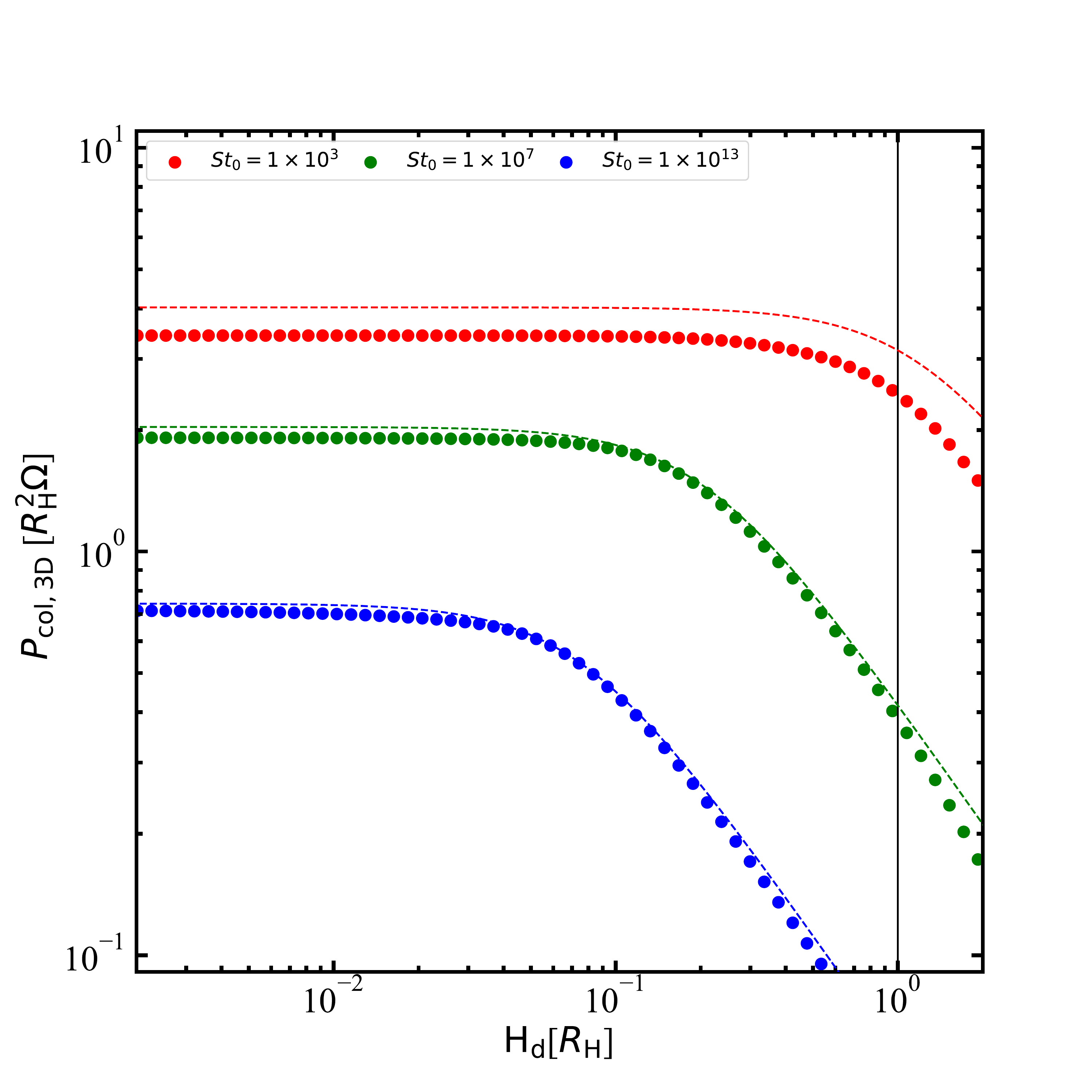}{0.33\textwidth}{(b) $P_{\rm col,3D}(i^{*})$}
	 \fig{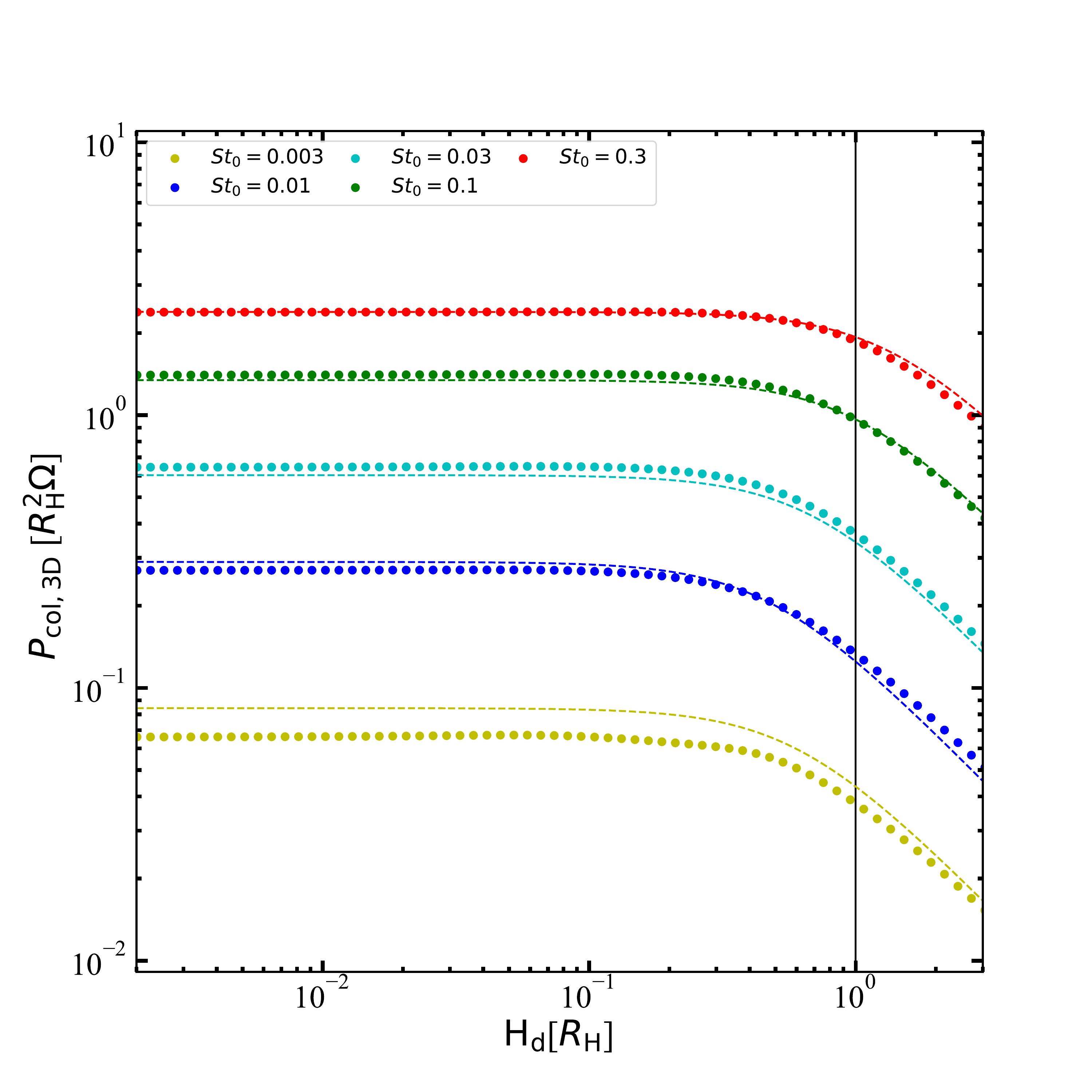}{0.33\textwidth}{(c) $P_{\rm col,3D}(H_{\rm d})$}
	 }
\caption{The same as Figs. \ref{fig:pcol2dm01}, \ref{fig:pcol3dlargem01}, and \ref{fig:pcol3dsmallm01}, but for $m=0.05$.}\label{fig:m005}
\end{figure*}

\begin{figure*}[htb!]
	 \gridline{
	 \fig{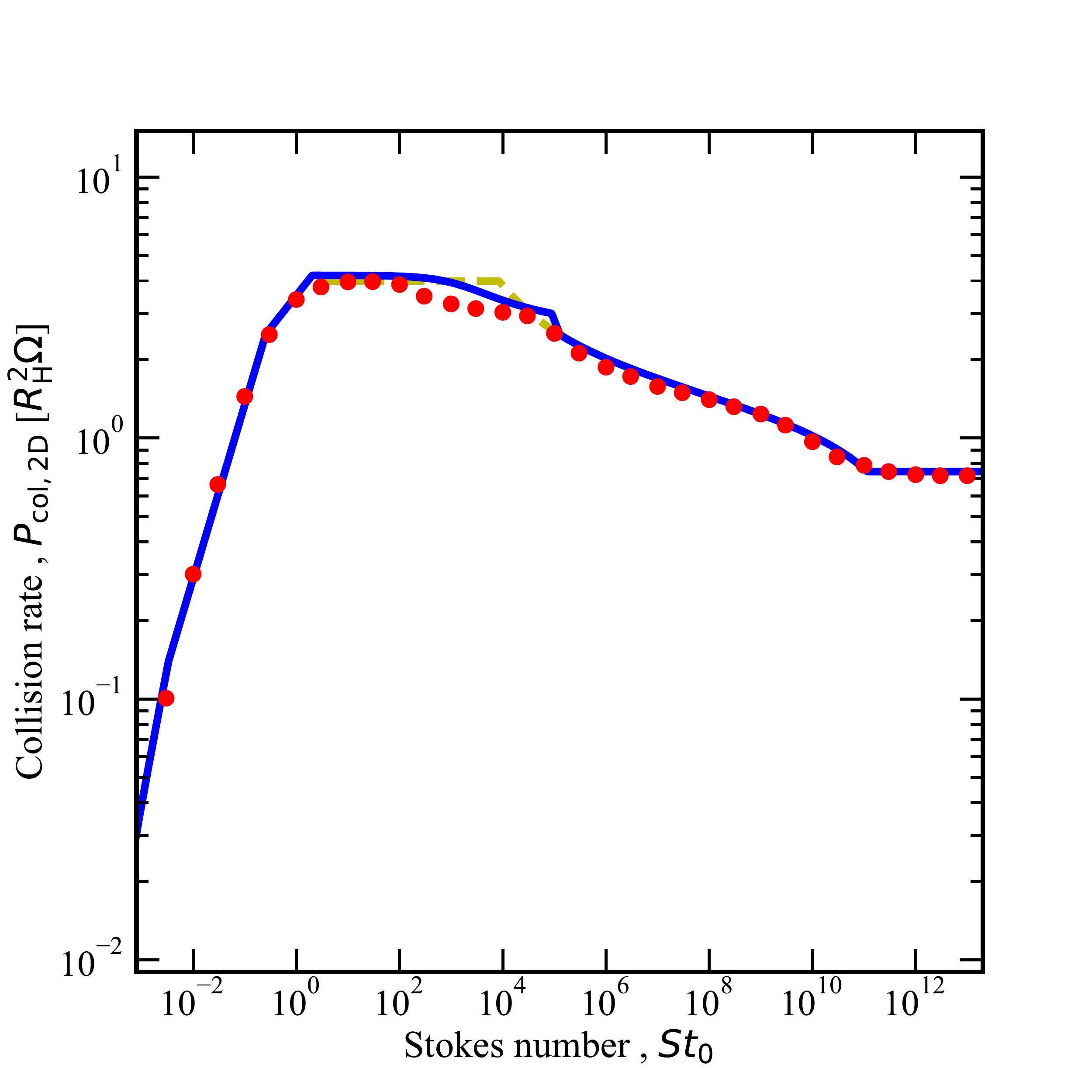}{0.33\textwidth}{(a) $P_{\rm col,2D}$}
	 \fig{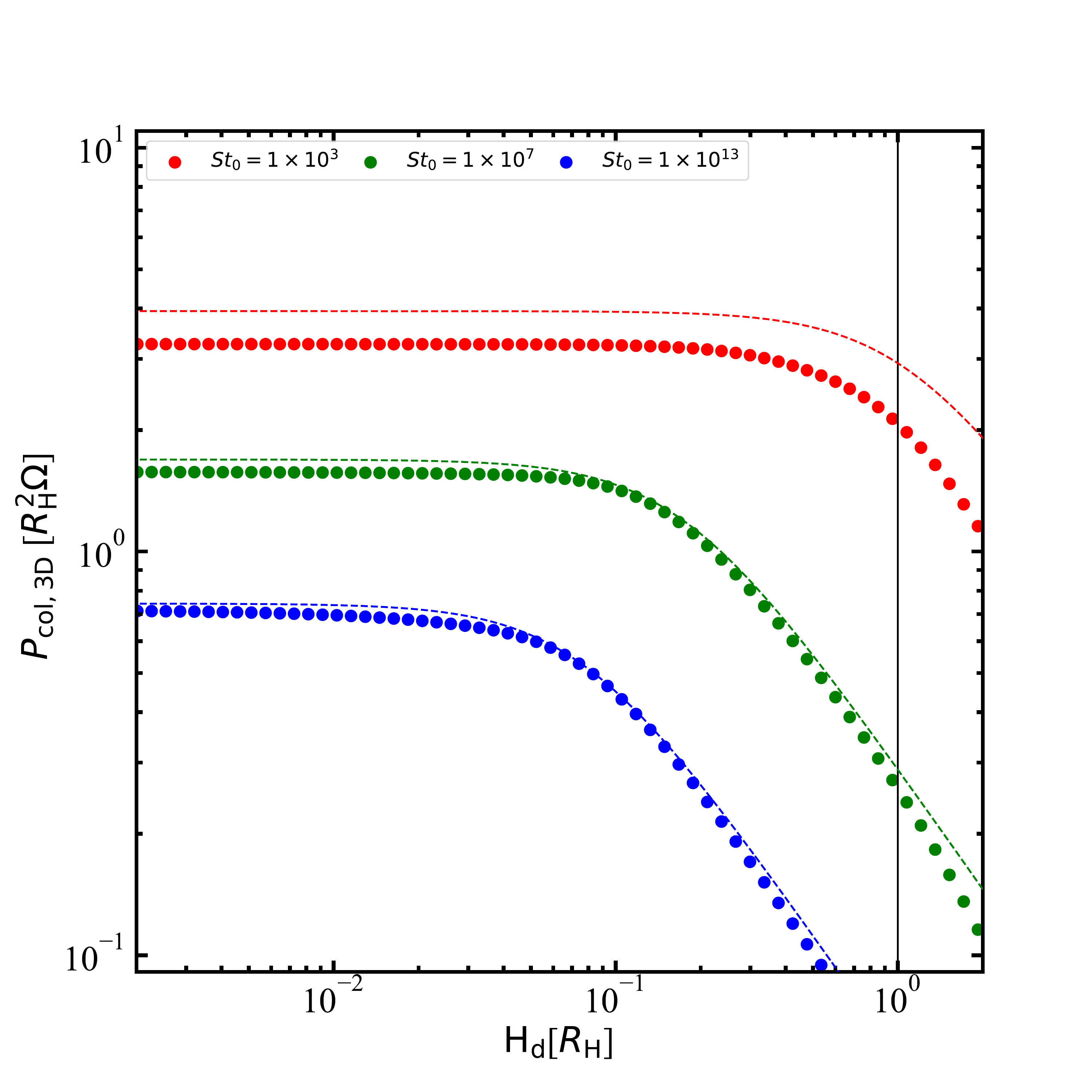}{0.33\textwidth}{(b) $P_{\rm col,3D}(i^{*})$}
	 \fig{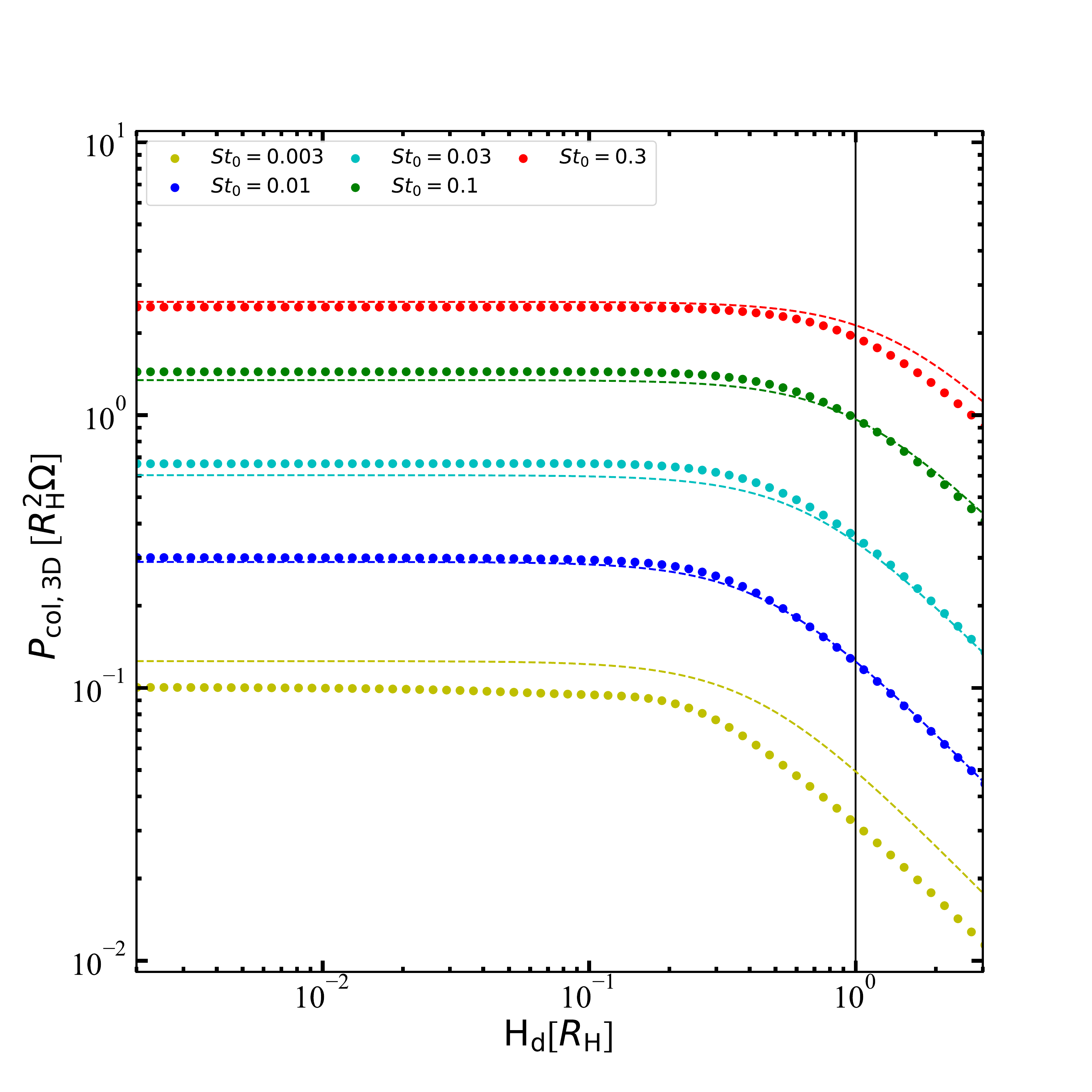}{0.33\textwidth}{(c) $P_{\rm col,3D}(H_{\rm d})$}
	 }
\caption{The same as Figs. \ref{fig:pcol2dm01}, \ref{fig:pcol3dlargem01}, and \ref{fig:pcol3dsmallm01}, but for $m=0.03$.}\label{fig:m003}
\end{figure*}

In the following equations, we summarize the two dimensional collision rate.
\begin{equation}
P_{\rm col,2D} = {\rm MIN}(P_{\rm col,atm},\ P_{\rm col,ss},\ P_{\rm col,set}, \ P_{\rm col,ho}),
\end{equation}
where
\begin{eqnarray}
P_{\mathrm{col},\mathrm{atm}} &=& P_{\rm col,free}(R_{\rm cap}), \label{eq:pcolatmsum} \\ 
\frac{P_{\mathrm{col},\mathrm{ss}}}{R_\mathrm{H}^2\Omega} &=& 2\sqrt{6} C_1\left(\frac{\rho_\mathrm{p}}{\rho_\mathrm{g}}\frac{c_\mathrm{s}}{R_\mathrm{H}\Omega}\frac{l_\mathrm{mfp}}{R_\mathrm{H}}St_0\right)^{1/4}, \label{eq:pcolsssum} \\ 
\frac{P_{\mathrm{col},\mathrm{set}}}{R_\mathrm{H}^2\Omega} &=& 3\left(2C_1 St_0\right)^{2/3},  \label{eq:pcolsetsum} \\
\frac{P_{\mathrm{col},\mathrm{ho}}}{R_\mathrm{H}^2 \Omega} &=& 2\frac{R_\mathrm{B}}{R_\mathrm{H}}\left[\frac{3 St_0}{\left(R_\mathrm{B}/R_\mathrm{H}\right)^2}-\xi\sqrt{\frac{3}{\left(R_\mathrm{B}/R_\mathrm{H}\right)}}\right], \nonumber \\ \label{eq:pcolhosum}
\end{eqnarray}
$P_{\rm col,free}(R_{\rm cap})$ is given in Eq. (\ref{eq:pcolfree2}).
Figures \ref{fig:pcol2dm01}, \ref{fig:m005}, and \ref{fig:m003} show the analytical solution and simulation results for $m=0.1$, $0.05$, and $0.03$, respectively. Our analytic formulae are  in agreement with our simulation.

If you feel the formula for $P_{\rm col,free}$ in $P_{\rm col,atm}$ is complicated, $P_{\rm col,atm}$ can be simple by setting the planetary radius $R_{\rm pl}$ to the capture radius $R_{\rm cap}$ in Eq. (\ref{eq:pcolfree}), thus
\begin{equation}
\frac{P_{\mathrm{col},\mathrm{atm}}}{R_\mathrm{H}^2\Omega} =11.3\sqrt{{\rm MIN}(R_{\rm cap}/R_{\rm H},1/8)}.
\label{eq:simplepcol}
\end{equation}
Although the maximum of $R_{\rm cap}$ is much larger than $R_{\rm H}/8$, the simple formula overestimates $P_{\rm col}$ for large $R_{\rm cap}$ so that we give the upper limit $R_{\rm cap}=R_{\rm H}/8$ in Eq. (\ref{eq:simplepcol}).
The simple formula is shown by the yellow dashed lines in Figs. \ref{fig:pcol2dm01}, \ref{fig:m005}, and \ref{fig:m003} and almost same as the more accurate formula.

If the vertical distribution is wide enough ($H_{\rm d} \gtrsim R_{\rm H}$), we need the three-dimensional specific collision rate $P_{\rm col,3D}$.
For $St_0>1$, $P_{\rm col,2D} \approx P_{\rm col,atm}$ and the vertical distribution of particles is determined by orbital inclinations $i$.
For $i^{*}$, the dispersion of $i$, $P_{\rm col,3D}$ is given by
\begin{eqnarray}
\frac{P_\mathrm{col,3D}(i^*)}{R_\mathrm{H}^2\Omega} &=&  \left\{\left(\frac{P_{\rm col,2D}}{R_\mathrm{H}^2\Omega}\right)^{-2} \right.  \nonumber \\
&&\left. +\left[\frac{R_{\rm cap}^2}{4\pi a i^* R_\mathrm{H}}\left(17.3+\frac{232 R_\mathrm{H}}{R_{\rm cap}}\right)\right]^{-2}\right\}^{-1/2}, 
\label{eq:3danalargesum}
\end{eqnarray}
where $i^{*}$ is related to the scale hight $H_{\rm d}$ as $i^{*}=\sqrt{2} H_{\rm d}/a$ (See Eq. \ref{eq:hd_iasta}).
On the other hand, for $St_0<1$, $P_{\rm col}$ is determined by $P_{\rm col,ss}$, $P_{\rm col,set}$, or $P_{\rm col,ho}$ and $P_{\rm col,3D}(H_{\rm d})$ is given by
\begin{equation}
P_\mathrm{col,3D}(H_{\rm d})= \left[\left( P_{\mathrm{col,2D}}\right)^{-2}+\left(P_{\mathrm{col,2D}}\frac{x_{\rm ss}}{0.65H_\mathrm{d}}\right)^{-2}\right]^{-1/2},
\label{eq:3danasmallsum}
\end{equation}
where $x_{\rm ss}$ is given by Eq. (\ref{eq:xss}) if $P_{\rm col,2D}=P_{\rm col,ss}$, $x_{\rm ss}$ is given by Eq. (\ref{eq:xset}) if $P_{\rm col,2D}=P_{\rm col,set}$, and $x_{\rm ss}=2 R_{\rm B}$ if $P_{\rm col,2D}=P_{\rm col,ho}$.

We compare the analytic formula for $m=0.1$, $0.05$, and $0.03$ in Figs. \ref{fig:pcol3dlargem01}, \ref{fig:pcol3dsmallm01}, \ref{fig:m005}, and \ref{fig:m003}.
The formula is in agreement with simulations for $m=0.03-0.1$.

Table \ref{tab:sumana} shows the summary of how the collision rate $P_{\rm col}$ can be obtained.

\begin{table*}[htbp]
	\begin{center}
	\caption{Summary of calculating the analytic collision rate.
}
	\begin{tabular}{lll}
	 \hline \hline 
	 $P_{\rm col, 2D}$\\
	1. Determine the size or stopping time of a particle: & Eq. (\ref{eq:stradii}) &  \\
	2. Calculate the capture radius $R_{\rm cap}$: & Eqs. (\ref{eq:dele}) and (\ref{eq:rcap}) & \\
	3. Calculate $P_{\rm col,2D}$ in different regimes :& $P_{\rm col,atm}$$\quad$Eq. (\ref{eq:pcolatmsum}) or Eq. (\ref{eq:simplepcol})&  \\
	& $P_{\rm col,ss}$$\quad$Eq. (\ref{eq:pcolsssum})\\
	& $P_{\rm col,set}$$\quad$Eq. (\ref{eq:pcolsetsum})\\
	& $P_{\rm col,ho}$$\quad$Eq. (\ref{eq:pcolhosum})\\
	4. Result of  $P_{\rm col,2D}$:  & $P_{\rm col,2D}=$ MIN($P_{\rm col,atm}$, $P_{\rm col,ss}$, $P_{\rm col,set}$, $P_{\rm col,ho}$) & \\ \hline
	 $P_{\rm col,3D}$ \\
	 5. Determine the distribution of particles: & $i^{*}\ {\rm or}\ H_{\rm d}$ ($i^{*}=\sqrt{2}H_{\rm d}/a$)\\
	6. Calculate the impact parameter $x_{\rm ss}$ :& Eq. (\ref{eq:xss}) if $P_{\rm col,2D}=P_{\rm col,ss}$ \\
	& Eq. (\ref{eq:xset}) if $P_{\rm col,2D}=P_{\rm col,set}$ \\
	& $x_{\rm ss}=2R_{\rm B}$ if $P_{\rm col} = P_{\rm col,ho}$ \\
	7. Result of $P_{\rm col ,3D}$: & Eq. (\ref{eq:3danalargesum}) if $P_{\rm col,2D}=P_{\rm col,atm}$ \\
	& Eq. (\ref{eq:3danasmallsum}) otherwise
    	\end{tabular}
	\label{tab:sumana}
	\end{center}
\end{table*}

\section{Discussion}\label{sec:dis}
\subsection{Comparison with Previous Studies}
We derive the analytic solutions for the collision rates.
These formulae are improved comparing to the previous studies, as explained below.

Large particles are captured via planetary atmosphere.
Protoplanets larger than the Moon can have an atmosphere \citep[e.g.,][]{mizuno1978}. 
\citet{inaba2003} obtained the analytic formula for $P_{\rm col}$ using the capture radius $R_{\rm cap}$ instead of the planetary radius, $R_{\rm pl}$.
\rev{They set the atmospheric outer boundary is given by $R_{\rm B}$ if $R_{\rm B} < R_{\rm H}$.}
\rev{However, the hydrodynamic simulation shows the atmospheric radius, inside which the density is enhanced, is given by the Hill radius rather than by the Bondi radius (see Fig. \ref{fig:rhoanam01}), resulting in $R_{\rm cap} \sim R_{\rm H}$ for small particles.}
However, their formula overestimates $P_{\rm col}$ for $R_{\rm cap}\sim R_{\rm H}$, because Eq. (\ref{eq:pcolfree}) is valid only for $R_{\rm cap} \ll R_{\rm H}$.
As discussed in \S \ref{sec:pcol2datm}, we thus improve the formula for $R_{\rm cap} \sim R_{\rm H}$.
On the other hand, \citet{inaba2003} considered a single encounter.
Multiple encounters are important for $R_{\rm cap} \sim R_{\rm pl}$ so that we take into account the effect (see Eq. \ref{eq:dele}).

We consider the super-sonic gas drag as well as the Stokes gas drag.
In the previous studies for $St_0<1$, mainly the Stokes or Epstein gas drag law were considered for discussing the pebble accretion.
However, the velocity of the particles can be greater than the sound velocity, so that the super-sonic gas drag is effective (see Eq. \ref{eq:velterm}).
We derive the analytic solution for $St_0 \sim 1$ considering the super-sonic gas drag.
This solution enables smooth connection of analytic solutions, between the atmospheric and settling regimes.

For the even smaller particles ($St_0 \ll 1$), the flow pattern of the gas due to horseshoe and Keplerian shear limits the accretion of particles onto the planet, as pointed out in the previous studies \citep{popovas2018, kuwahara2020a, kuwahara2020b, homma2020}.
We derive the analytic solution, considering the flow effect.
For the three-dimensional case, the vertical gas flow enters from high latitudes.
Thus, particles can accrete onto the planet from high latitudes.
This three-dimensional effect is also shown in the previous studies.
Taking into account the effect, we derive the analytical formula for the vertical distribution dependence of $P_{\rm col}$ (see \S \ref{sec:pcol3d}).

\subsection{Collision Rate for Realistic Atmosphere}\label{sec:realden}
In our simulation, the density of the atmosphere is unrealistic because the density profile reaches the hydrostatic isothermal solution and becomes shallower in the vicinity of the planet due to the softening.
As showed above, the density profile is consistent with the hydrostatic equilibrium in $r \lesssim R_{\rm H}$ (see Fig. \ref{fig:rhoanam01}).
Therefore, adopting the hydrostatic density profile of the atmosphere according to the accretion heating and the opacity, we then obtain realistic $P_{\rm col}$.
\rev{In this subsection, we calculate the analytic collision rate using the more realistic density profile for the atmpsphere than that obtained by the simulation.}

\rev{We consider low opacity atmospheres, where the energy transportation is dominated by radiation rather than convection. For $r \ll R_{\rm B}$, the analytical atmospheric density profile is approximately given by \citep{inaba2003, kobayashi2011}}
\begin{equation}
\rho_{\rm atm}=\frac{\pi \sigma_{\rm SB}}{12 \kappa L_{\rm e}}\left(\frac{G M_{\rm p} \mu m_{\rm H}}{k_{\rm B}} \right)^4\frac{1}{r^3},
\label{eq:realden}
\end{equation}
where $\kappa$ is the opacity of the atmosphere, $k_{\rm B}$ is the Boltzmann constant, and $\sigma_{\rm SB}$ is the Stefan-Boltzmann constant.
The planetary luminosity $L_{\rm e}$ mainly comes from the accretion of bodies. We assume
\begin{equation}
L_{\rm e}=\frac{GM_{\rm p}}{R_{\rm pl}}\frac{dM_{\rm p}}{dt}.
\label{eq:luminosity}
\end{equation}
We calculate $R_{\rm cap}$ according to the density profile given in Eq. (\ref{eq:realden}) and then obtain $P_{\rm col}$ from our analytic formula.

\begin{figure}[htb!]
	\centering
	\includegraphics[scale=0.32]{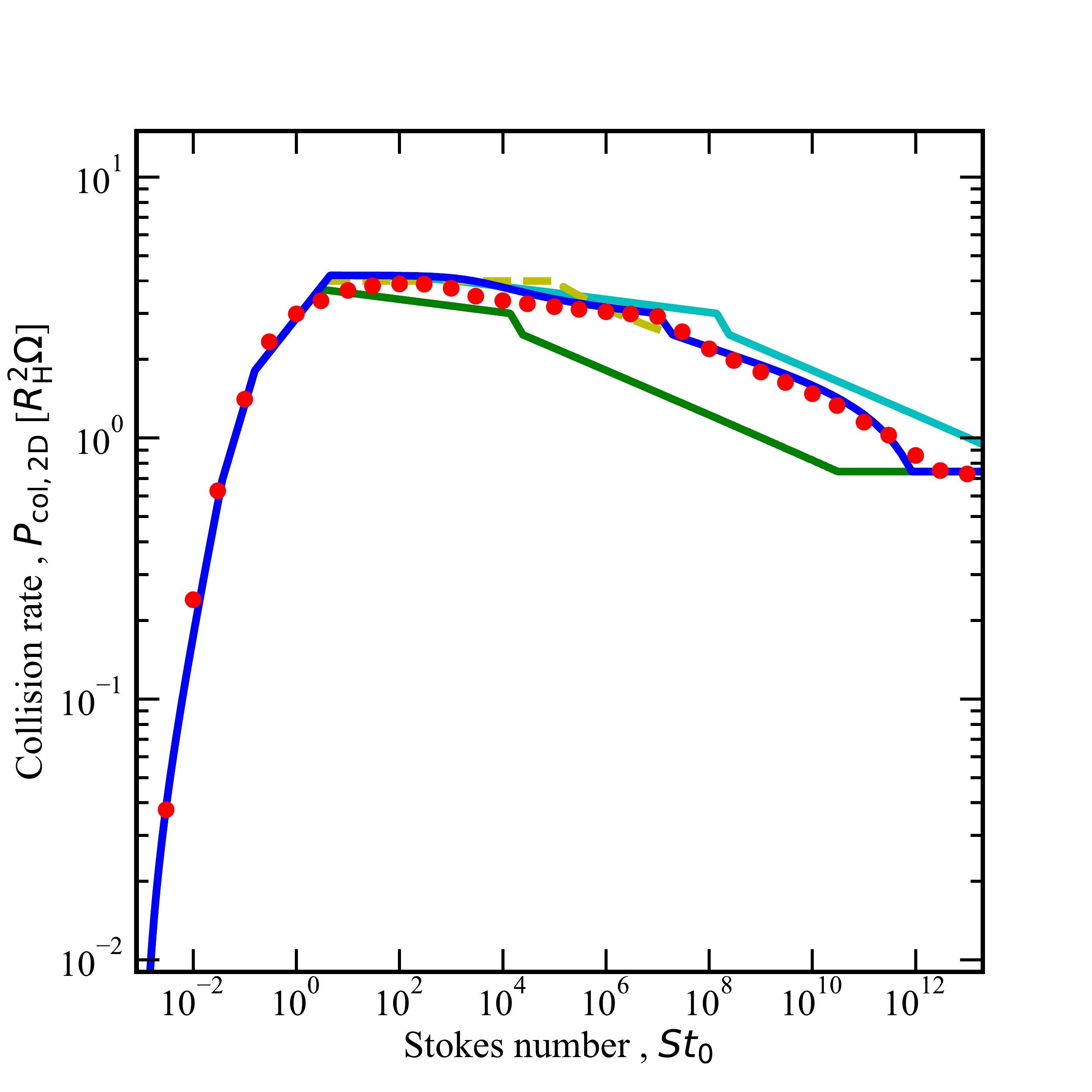}
	\caption{Same as Fig. \ref{fig:pcol2dm01} but for the case of the more realistic atmosphere. The green and cyan lines correspond to the analytical solutions (Table \ref{tab:sumana}) for $\kappa=1\ {\rm cm^2\ g^{-1}}$ and $\kappa=0.01\ {\rm cm^2\ g^{-1}}$.}
	\label{fig:2dpcolm01real}
\end{figure}

Figure \ref{fig:2dpcolm01real} shows the analytical two-dimensional specific collision rates (Table \ref{tab:sumana}) for the more realistic density profile, assuming $d M_{\rm p} / dt = 1 \times 10^{-6} M_{\oplus} {\rm yr}^{-1} $.
The density profile in Eq. (\ref{eq:realden}) with $\kappa=0.01\ {\rm cm}^2\ {\rm g}^{-1}$ is similar to the Eq. (\ref{eq:denana}) so that the collision rate is similar to our simulation (Fig. \ref{fig:2dpcolm01real}). For the planet formation, the accretion rate and $P_{\rm col}$ are determined consistently \citep[e.g.,][]{inabaet2003, chambers2006, kobayashi2011,kobayashi2018}.

\subsection{Estimate of Planetary Growth}
In this subsection, we estimate the accretion timescale $T_{\rm acc}$ and the required disk mass $M_{\rm req}$ for the formation of a planet with mass $M_{\rm p}$.
Solid materials drift inward to the protoplanetary disk. The drift velocity $v_{\rm drift}$ is given by \citep{weidenschilling1977a, adachi1976}
\begin{equation}
v_{\rm drift} = -\frac{2 St_0}{1+St_0^{2}}v_{\rm hw},
\end{equation}
where $v_{\rm hw}$ is the headwind velocity.
We define the accretion efficiency of solids
\begin{equation}
\varepsilon \equiv \frac{\dot{M}_{\rm p}}{\dot{M}_{\rm drift}} = \frac{P_{\rm col}\Sigma_{\rm d}}{2 \pi a \Sigma_{\rm d}|v_{\rm drift}|},
\end{equation}
where $\Sigma_{\rm d}$ is the surface density of solids.
The required disk mass and the accretion timescale are, respectively, defined by
\begin{equation}
M_{\rm req} \equiv \frac{\chi M_{\rm p}}{\varepsilon},
\end{equation}
\begin{equation}
T_{\rm acc} \equiv \frac{M_{\rm p}}{\dot{M}_{\rm p}}.
\end{equation}
where $\chi$ is the gas to solid ratio.
We assume the particle scale hight is given by \citep{youdin2007}
\begin{equation}
H_{\rm d} = \left(1+\frac{St_0}{\alpha}\frac{1+2 St_0}{1+St_0}\right)^{-1/2} H,
\end{equation}
where $\alpha$ is the turbulent parameter \citep{shakura1973}.

\begin{figure}[htb!]
	\centering
	\includegraphics[scale=0.28]{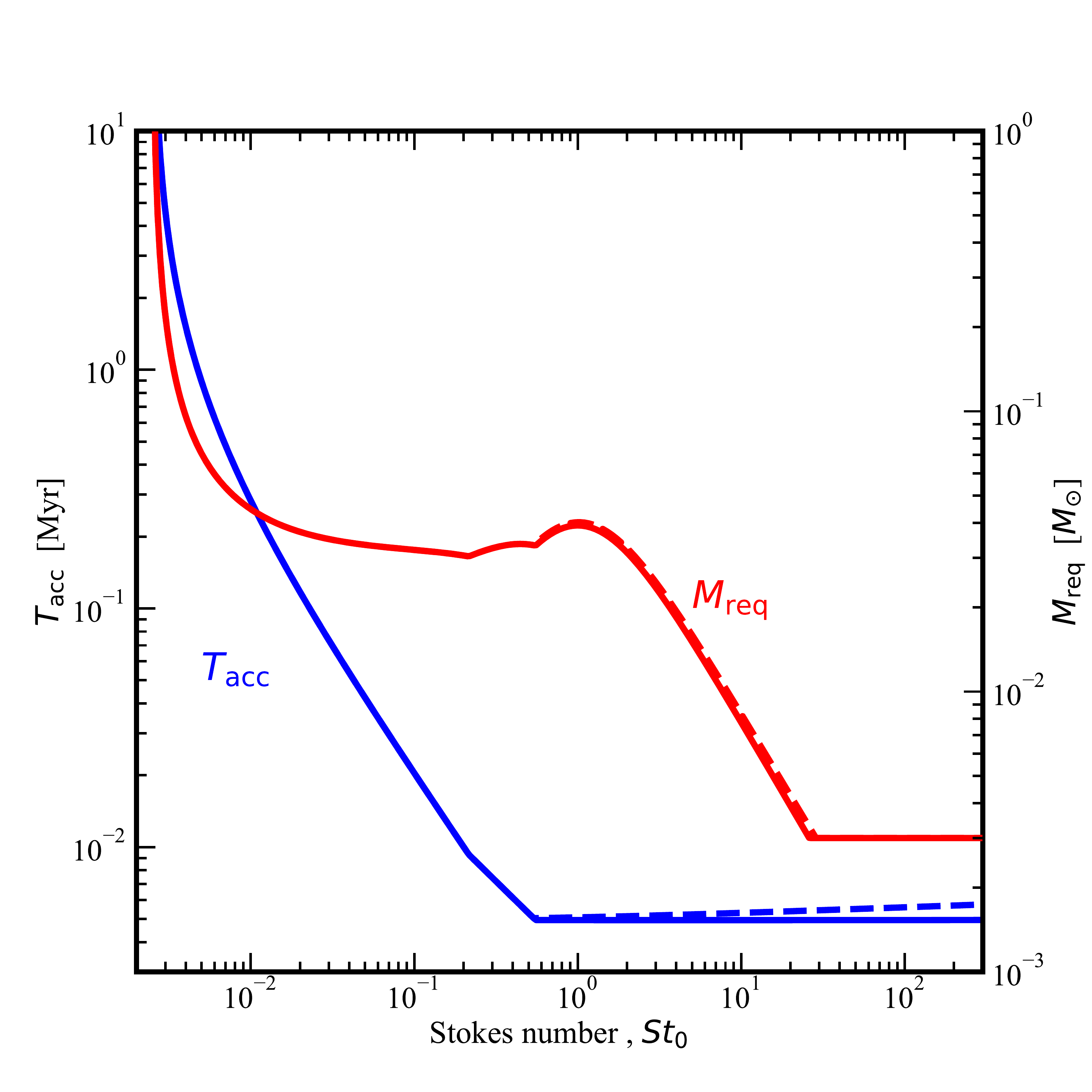}
	\caption{Blue and red lines show the accretion timescale and the required disk mass for $\kappa=0.01{\rm cm^2\ g^{-1}}$ (solid), $1{\rm cm^2\ g^{-1}}$ (dased). Note that our analytic formulae are applicable for $ St_0 \la 100$ (see Appendix \ref{sec:appC}).}
	\label{fig:estimation}
\end{figure}

Figure \ref{fig:estimation} shows $T_{\rm acc}$ and $M_{\rm req}$ for a planetary mass $M_{\rm p}=10M_{\oplus}$, $a=5\ {\rm au}$, $\Sigma_{\rm d}=2\ {\rm g\ cm^{-2}}$, $v_{\rm hw} \approx 50\ {\rm m\ s^{-1}}$, $\chi =100$, and $\alpha=1\times10^{-3}$. We obtain $P_{\rm col}$ via the analytic formula shown in Table \ref{tab:sumana} for low eccentricity partciles. Therefore our estimate is valid for $St_0 \lesssim 100$ (see Appendix \ref{sec:appC})

Very small particles ($St_0 \lesssim 10^{-3}$) are hardly accreted onto the planet (see \S \ref{sec:hoeffect}), so that the accretion timescale and required mass are huge.

For $St_0 \sim 1$, the accretion timescale is much shorter than the disk lifetime as the previous studies have shown \citep[e.g,][]{ormel2010}. However, a massive disk is required for the formation of the single core of a gas giant planet.

For large particles ($St_0 \gtrsim 10$), the accretion timescale is as short as that for $St_0 \sim 1$, because of the atmospheric enhancement. The accretion efficiency is close to the unity because the drift velocity is slow.  
The required disk mass is thus a constant value; $M_{\rm req} \approx \chi M_{\rm p}$ because of $\varepsilon \approx 1$.

The radial drift of particles with $St_0 \sim 1$ effectively supplies materials for planet growth in the inner disk.
However, the accretion efficiency of such particles is low.
Once collisional growth among drifting particles results in large bodies with $St_0 \gtrsim 10$, the cores of giant planets are effectively formed. We will address this issue by treating the collisional evolution and radial drift consistently \cite[e.g.,][]{kobayashi2018}.

\section{Summary and Conclusions}\label{sec:con}
In this paper, we investigate the effect of the protoplanetary disk perturbed by the planet on the collision rate of particles with the planet. We perform the non-isothermal three dimensional hydrodynamic simulation in the frame co-rotating with the planet and then integrate the equation of motion of particles in the gas flow obtained from the simulation considering the super-sonic and Stokes gas drag.
We then derive the new analytic formulae for the collision rate, considering the following regimes:
\\

1. Meter-sized or larger particles are captured via the planetary atmosphere. Because they approach the planet exceeding the sound velocity, they feel strong gas drag. The atmosphere decelerates and captures the particles. The collision rate is significantly enhanced by the atmosphere.
\\

2. Smaller particles are influenced by the gas flow in the protoplanetary disk effectively. They are well coupled to the gas flow. The collision rates are determined by comparing two timescales: the encounter timescale in which a particle has a close encounter with a planet and the settling timescale in which a particle drifts onto a planet. If the settling timescale is shorter than the encounter timescale, a particle collide with a planet.
For relatively large particles, the drift velocity is determined by the super-sonic gas drag, while the Stokes gas drag is dominant for smaller particles.
\\

3. If the particles are very small, the horseshoe gas flow and the outflow around the planet prevent particles from colliding with the planet. As a result, the collision rate sharply decreases with the decreasing size of particles. Particles can collide with the planet in the narrow band between the Keplerian shear and horseshoe flows.
\\

These analytical formulae (summarized in Table \ref{tab:sumana}) are in good agreement with our numerical simulations.
In the three-dimensional case, we also derive the analytic formulae (Eqs. \ref{eq:3danalargesum} and \ref{eq:3danasmallsum}) and confirm the consistency between simulations and the analytical solutions.
We show the method to obtain the analytic collision rate with the analytic formulae in Table \ref{tab:sumana} and \S\ref{sec:summarypcol}.

We estimate the formation timescale of a solid core for the gas giant formation (Fig. \ref{fig:estimation}). The formation timescale is much shorter than the disk lifetime if $St_0=10^{-2}$ to $10^{2}$.
However, the drift velocity is so high that the accretion efficiency is small for $St_0 \lesssim 10$.
Therefore, the collisional evolution between pebbles drifting from the outer disk may be important to reconcile the issue.
Our results are helpful to discuss the planet formation in a wide size distribution of bodies.

\begin{acknowledgments}
We are grateful to the anonymous referee for helpful comments, which significantly improves the original version of our manuscript. 
We would like to thank Elijah Mullens for valuable comments. 
This work is supported by the financial support of JSPS KAKENHI Grant (17K05632, 17H01103, 17H01105, 18H05438, 18H05436, 20H04612, 21K03642). Hydrodynamic simulations in this work were carried out on the Cray XC50 supercomputer at the Center for Computational Astrophysics, National Astronomical Observatory of Japan. We thank Athena++ developers: James M. Stone, Kengo Tomida, Christopher White, and Kyle Gerard Felker.
\end{acknowledgments}
\appendix
\section{Analytical Solution of the Density Profile}\label{sec:appA}
In our hydrodynamic simulations, we adopt the $\beta$ cooling model. The density profile reaches the isothermal solution in the quasi-steady state.
To solve the density profile analytically, we assume the hydrostatic equilibrium in the isothermal case.
The equation for the hydrostatic equilibrium is given by
\begin{equation}
\frac{\partial p}{\partial r}=-\frac{G M_\mathrm{p}r}{\left( r^2+r_\mathrm{s}^2\right)^{3/2}}\rho_\mathrm{g}.
\end{equation}
We use the relation between the density and the pressure in the isothermal, $p=\rho_\mathrm{g} c_\mathrm{s}^2$.
We then have
\begin{equation}
\frac{1}{p}\frac{\partial p}{\partial r} = -\frac{R_\mathrm{B} r}{\left( r^2+r_\mathrm{s}^2\right)^{3/2}}.
\end{equation}
The solution to this equation is given by
\begin{equation}
p=p_0\ \mathrm{exp}\left(\frac{R_\mathrm{B}}{\sqrt{r^2+r_\mathrm{s}^2}}-\frac{R_\mathrm{B}}{\sqrt{R_\mathrm{H}^2+r_\mathrm{s}^2}}\right),
\end{equation}
\begin{equation}
\rho_{\rm g}=\rho_0\ \mathrm{exp}\left(\frac{R_\mathrm{B}}{\sqrt{r^2+r_\mathrm{s}^2}}-\frac{R_\mathrm{B}}{\sqrt{R_\mathrm{H}^2+r_\mathrm{s}^2}}\right),
\label{eq:denana}
\end{equation}
where we assume $p=p_0,\ \rho_{\rm g}=\rho_0$ at the Hill radius.

The solution is plotted in Fig. \ref{fig:rhoanam01} and in agreement with our simulation.


\restartappendixnumbering

\section{Relation between the Dust Scale Hight and the Dispersion of Inclinations} \label{sec:appB}
The large particles have orbital inclinations, so that the $z$-coordinate is given by
\begin{equation}
z = ia\ {\rm sin}\theta_{\rm s},
\end{equation}
where $\theta_{\rm s}$ is the true anomaly of the particle. Assuming the isotropic distribution of the true anomaly, the distribution function of $z$ for an arbitrary $i$, $f(z)$ is given by
\begin{equation}
f(z)=\frac{1}{\pi}\frac{1}{dz/d\theta_{\rm s}}=\frac{1}{\pi ia\ {\rm cos}\theta_{\rm s}}=\frac{1}{\pi i a\left(1-z^2/i^2a^2\right)^{1/2}}.
\end{equation}
The Rayleigh-type distribution function of inclination is given by
\begin{equation}
n(i)=\frac{2i}{i^{*2}}\mathrm{exp}\left(-\frac{i^2}{i^{*2}}\right).
\end{equation}
Thus, the distribution function of $z$ is given by
\begin{eqnarray}
\int^{\infty}_{z/a}f(z) n(i) di &=& \int^{\infty}_{z/a}\frac{2\ \mathrm{exp}\left(-i^{2}/i^{*2}\right)}{a \pi i^{*2} \left(1-z^2/i^2a^2\right)^{1/2}}di \nonumber \\
&=& \frac{1}{a \sqrt{\pi}i^{*}}\ {\rm exp}\left(-\frac{z^2}{i^{*2}a^2}\right).
\label{eq:incdis}
\end{eqnarray}
Eq. (\ref{eq:incdis}) corresponds to the Gaussian distribution of $z$.
On the other hand, the Gaussian distribution function of $z$ with the scale hight $H_{\rm d}$ is given by
\begin{equation}
\frac{1}{\sqrt{2\pi}H_\mathrm{d}}\mathrm{exp}\left(-\frac{z^2}{2 H_\mathrm{d}^2}\right).
\label{eq:gaussdis}
\end{equation}
Comparing the two distribution functions in Eqs. (\ref{eq:incdis}) and (\ref{eq:gaussdis}), we obtain
\begin{equation}
H_\mathrm{d}=\frac{a i^{*}}{\sqrt{2}}.
\label{eq:hd_iasta}
\end{equation}

\restartappendixnumbering

\section{Condition for $e\la R_{\rm H}/a$\label{sec:appC}}
We here estimate the dispersion of eccentricity for small particles around a planet according to \citet{kobayashi2010,kobayashi2011}.
The eccentricity damping rate due to the viscous stirring is given by \citep{ohtsuki2002}
\begin{equation}
\frac{de^{*2}}{dt} = n_{\rm M} a^2 \left(\frac{R_{\rm H}}{a}\right)^4 \langle P_{\rm VS} \rangle \Omega,
\end{equation}
where $e^{*}$ is the dispersion of eccentricity, $\langle P_{\rm VS} \rangle$ is the dimensionless stirring rate, and $n_{\rm M}$ is the surface number density of protoplanets given by
\begin{equation}
n_{\rm M}=\frac{1}{2^{4/3} \pi \tilde{b} R_{\rm H}a},
\end{equation}
where $\tilde{b} \simeq 10$ is a factor of the orbital separation of protoplanets \citep{kokubo2002}.
On the other hand, $e^{*}$-damping rate due to the gas drag is given by
\begin{equation}
\frac{d e^{*2}}{dt} = -2\frac{e^{*2}}{St_0} \Omega.
\end{equation}
Gas drag effectively damps $e^{*}$ of small particles, so that we adopt $\langle P_{\rm VS} \rangle = 73$ is independent of $e^{*}$ for $e^{*}\ll R_{\rm H}/a$.
Thus, the equilibrium condition between the stirring and the damping gives
\begin{equation}
e^{*2} = \frac{(R_{\rm H}/a)^3 \langle P_{\rm VS} \rangle \Omega St_0}{2^{7/3}\pi \tilde{b}}\sim 0.46 \left(\frac{R_{\rm H}}{a}\right)^3 St_0.
\end{equation}
For the planet formation with $M_{\rm p}=10M_{\oplus}$ ($R_{\rm H}/a\approx 0.02$), $e^{*}\la R_{\rm H}/a$ for $St_0 \la 100$. Therefore, our estimate in Fig. \ref{fig:estimation} is valid for $St_0 \la100$.

\bibliography{myref}
\end{document}